\documentclass[twocolumn,aps,prd,superscriptaddress,nofootinbib,floatfix,preprintnumbers]{revtex4-1}

\usepackage{graphicx,multirow}
\usepackage{xspace}

\usepackage{amsmath,amsfonts}

\usepackage{hyperref}
\usepackage{soul}

\newcommand{\nn}{\nonumber}
\newcommand{\beq}{\begin{equation}}
\newcommand{\eeq}{\end{equation}}
\newcommand{\beqa}{\begin{eqnarray}}
\newcommand{\eeqa}{\end{eqnarray}}

\newcommand{\babar}{\mbox{\ensuremath{{\displaystyle B}\!{\scriptstyle A}{\displaystyle B}\!{\scriptstyle AR}}}\xspace}

\newcommand{\Bbar}{\,\overline{\!B}{}}
\newcommand{\Dbar}{\,\overline{\!D}{}}
\newcommand{\Kbar}{\,\overline{\!K}{}}
\def\B0bar{\Bbar{}^0}
\def\D0bar{\Dbar{}^0}
\def\K0bar{\Kbar{}^0}

\begin{document}

\preprint{MITP/14-054}

\title{\boldmath New ways to search for right-handed current in 
$B\to \rho \ell \bar\nu$ decay}

\author{Florian U.\ Bernlochner}
\affiliation{University of Victoria, Victoria, British Columbia, Canada V8W 3P}
\affiliation{Physikalisches Institut der Rheinische Friedrich-Wilhelms-Universit\"at Bonn, 53115 Bonn, Germany}

\author{Zoltan Ligeti}
\affiliation{Ernest Orlando Lawrence Berkeley National Laboratory,
University of California, Berkeley, CA 94720}

\author{Sascha Turczyk}
\affiliation{PRISMA Cluster of Excellence \& Mainz Institut for Theoretical
Physics, Johannes Gutenberg University, 55099 Mainz, Germany}

\begin{abstract}

An interesting possibility to ease the tension between various determinations of
$|V_{ub}|$ is to allow a small right-handed contribution to the standard model
weak current.  The present bounds on such a contribution are fairly weak. We
propose new ways to search for such a beyond standard model contribution in
semileptonic $B\to\rho \ell \bar\nu$ decay.  Generalized asymmetries in one,
two, or three angular variables are introduced as discriminators, which do not
require an unbinned analysis of the fully differential distribution, and a
detailed study of the corresponding theoretical uncertainties is performed.   A
discussion on how binned measurements can access all the angular information
follows, which may be useful in both $B\to \rho \ell \bar\nu$ and $B\to K^*
\ell^+\ell^-$, and possibly essential in the former decay due to backgrounds. 
The achievable sensitivity from the available \babar and Belle data sets is
explored, as well as from the anticipated $50\,{\rm ab}^{-1}$ Belle~II data.

\end{abstract}

\maketitle

\section{Introduction}\label{sec:intro}

There is a long standing persistent tension between measurements of $|V_{ub}|$
from $B$ decays in leptonic, inclusive semileptonic, and exclusive semileptonic
decay channels.  In semileptonic decays, the difference between the inclusive
determination and that based on $B \to \pi \, \ell \bar\nu$ is almost
3$\sigma$.  It is possible that the resolution of this is related to not
sufficiently understood theoretical or experimental issues, and future theory
progress combined with the anticipated much larger Belle~II data sets will yield
better consistency.  A precise determination of $|V_{ub}|$ is crucial for
improving tests of the standard model (SM) and the sensitivity to new physics in
$B^0-\B0bar$ mixing~\cite{Charles:2013aka}. 

Another possibility, which received renewed attention
recently~\cite{Crivellin:2009sd, Buras:2010pz}, is that this tension can be
eased by allowing for a right-handed admixture to the SM weak current.  Such a
contribution could arise from not yet discovered TeV-scale new physics.  In
general, from a low energy effective theory point of view, the SM can be
extended by several new operators relevant for semileptonic decays, suppressed
by ${\cal O}(v^2/\Lambda^2)$~\cite{Feger:2010qc,Faller:2011nj}, where $\Lambda$
is a high scale related to new physics.  For simplicity, we consider the
effective Lagrangian with only one new parameter,
\begin{equation}\label{finalLNew}
\mathcal{L}_\text{eff} = -\frac{4 G_F}{\sqrt2} V_{ub}^L
  \big(\bar{u}\gamma_\mu P_L b 
  + \epsilon_R\, \bar{u}\gamma_\mu P_R b\big)
  (\bar\nu\gamma^\mu P_L \ell)+ \text{h.c.} ,
\end{equation}
where $P_{L,R} = (1\mp\gamma_5)/2$. The SM is recovered as $\epsilon_R\to 0$. 
Since we consider observables with leading, linear, dependence on
Re\,$(\epsilon_R)$, we assume it to be real in this paper, unless indicated
otherwise.  This happens to be the expectation in models with flavor structures
close to minimal flavor violation.  We do not consider $b\to c\ell\nu$ decay in
this paper, as the tension between the determinations of $|V_{cb}|$ is
less severe, and the connection between $b\to u$ and $b\to c$ transitions is
model dependent (see, however, Ref.~\cite{Crivellin:2014zpa}).
To distinguish from determinations of $|V_{ub}|$ assuming the SM, we refer to
analyses which allow for $\epsilon_R \neq 0$ as measurements of $|V_{ub}^L|$.

\begin{table}[b]
\begin{tabular}{ccc}
\hline\hline
Decay  &  $\left| V_{ub} \right| \times 10^3$ & $\epsilon_R$ dependence  \\
\hline
$B\to \pi \, \ell \bar \nu$ & $3.23 \pm 0.30$  & $1+ \epsilon_R$  \\
$B\to X_u \ell \bar\nu$  & $4.39 \pm 0.21$ & $\sqrt{1+ \epsilon_R^2}$  \\
$B\to \tau \, \bar \nu_\tau$  &$4.32 \pm 0.42$  & $1- \epsilon_R$  \\
\hline
Decay  &  \multicolumn{2}{c}{${\cal B} \times 10^4$}  \\
\hline
$B\to \rho \, \ell \bar\nu$ & 
  \multicolumn{2}{c}{$1.97 \pm 0.16$  ($q^2 < 12$\,GeV$^2$)} \\
$B\to \omega \, \ell \bar\nu$ & 
  \multicolumn{2}{c}{$0.61 \pm 0.11$ ($q^2< 12$\,GeV$^2$)} \\
\hline\hline
\end{tabular}
\caption{The $|V_{ub}|$ measurements~\cite{Amhis:2012bh} used in the fit shown
in Fig.~\ref{fig:epsr_chisqfit} and their dependence on $\epsilon_R$. The
branching fractions are taken from Ref.~\cite{Sibidanov:2013rkk}}
\label{tab:fit_simple}
\end{table}

\begin{figure*}[t]
\centerline{
\includegraphics[width=.9\columnwidth]{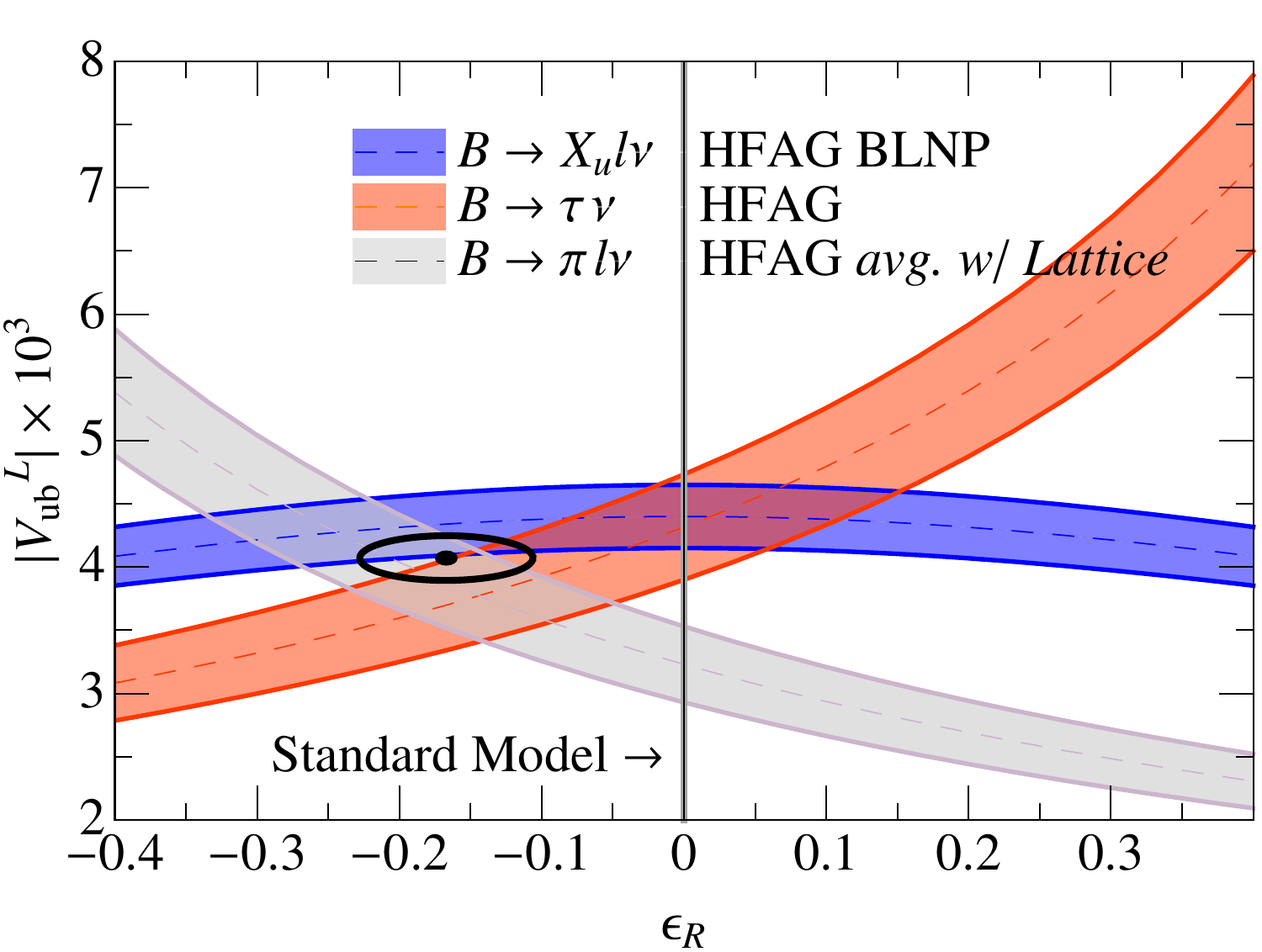}
\hfil
\includegraphics[width=.9\columnwidth]{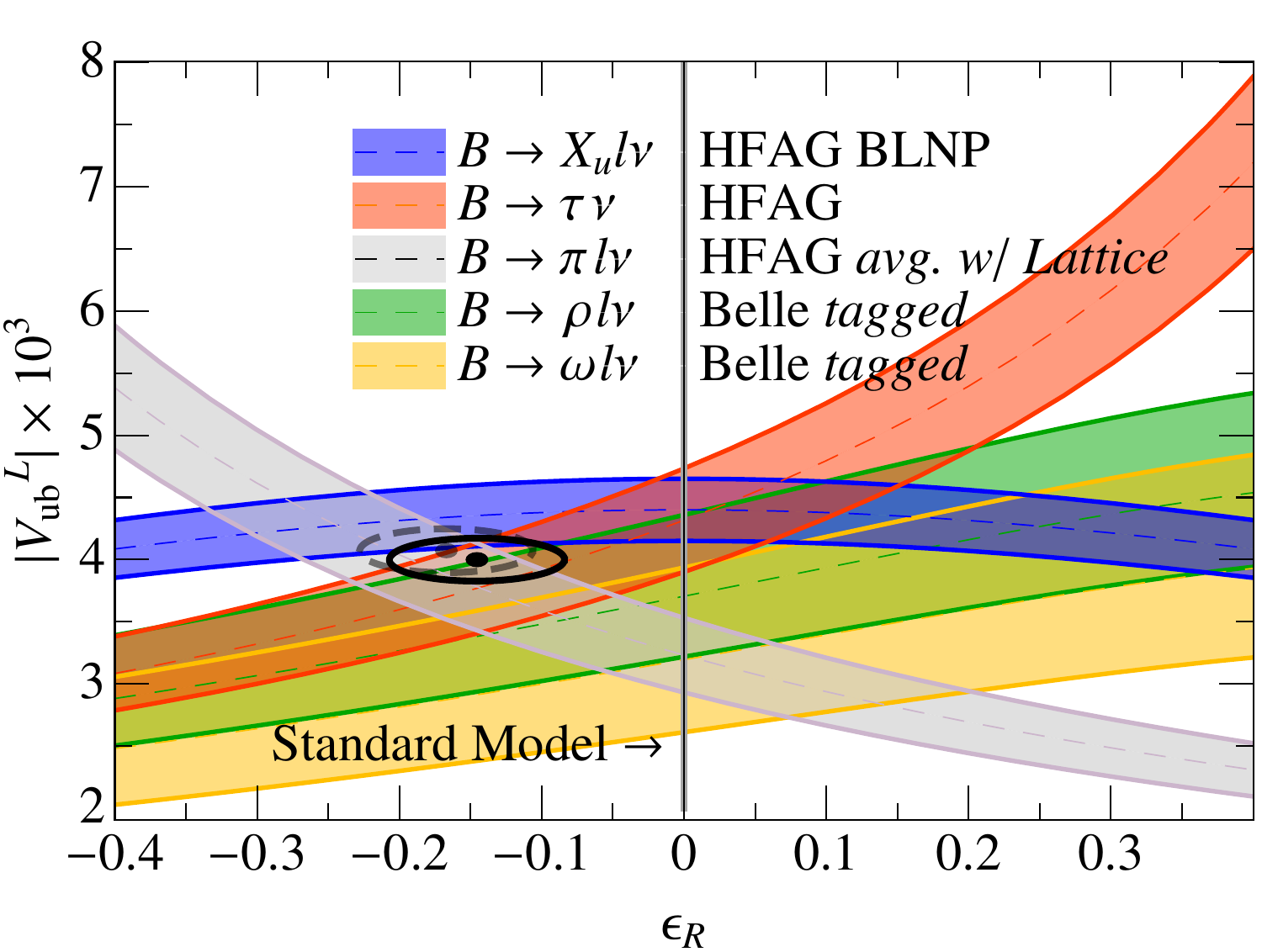}}
\caption{The allowed $|V_{ub}^L| - \epsilon_R$ regions. The black ellipse in the
left (right) plot shows the result of a $\chi^2$ fit using the first three
(four, excluding $\omega$) measurements in Table~\ref{tab:fit_simple}. The
fainter ellipse in the right plot is the same as that in the left plot.}
\label{fig:epsr_chisqfit}
\end{figure*}

The current measurements of $|V_{ub}|$ are summarized in
Table~\ref{tab:fit_simple}, and their dependence on $\epsilon_R$ is indicated in
the three cases in which it is simple.  The $\rho$ and $\omega$ measurements are
from Ref.~\cite{Sibidanov:2013rkk} using the theoretical predictions of
Ref.~\cite{Ball:2004rg}, and the two isospin-related $\rho$ modes were averaged
assuming a 35\% correlation of the systematic
uncertainties~\cite{Sibidanov:2013rkk}.  While we do not study the $\omega$
final state, it could provide complementary information in the future if lattice
QCD calculations of the form factors become available.  For $B \to X_u \ell
\bar\nu$ the BLNP result was used.   The result of the $\chi^2$ fit for
$|V_{ub}^L| - \epsilon_R$ without and with $B \to \rho\, \ell \bar\nu$ are shown
in Fig.~\ref{fig:epsr_chisqfit}.

The goal of this this paper is to devise observables sensitive to new physics
contributions in $\epsilon_R$, without requiring the measurement of the fully
differential decay distribution.  It is well-known from the literature on both
semileptonic and rare decays that a full description of the four-body final
state in $B \to \rho \ell \bar\nu$ depends on the dilepton invariant mass,
$q^2$, and three angles.  While we assume that the neutrino four-momentum is
reconstructed, past studies of $B\to D^*$~\cite{Aubert:2007rs, Dungel:2010uk}
and $D\to\rho$~\cite{CLEO:2011ab} semileptonic decays show that for $B \to \rho
\ell \bar\nu$, which has a much smaller rate, the full angular analysis will be
challenging and may be many years in the future.  Thus, it is interesting to
explore how the best sensitivity to $\epsilon_R$ may be obtained using current
and near future data sets.

\begin{table}[t]
\begin{tabular}{ccccc}
\hline\hline
Fit  &  $| V_{ub}^L| \times 10^4$ & $\epsilon_R$ & $\chi^2$\,/\,ndf & Prob. \\
\hline
3 modes  &  $4.07 \pm 0.18$ & $-0.17 \pm 0.06$  & 2.5\,/1 & 0.11   \\
4 modes  &  $4.00 \pm 0.17$ & $-0.15 \pm 0.06$   & 4.5\,/2 & 0.11 \\
\hline\hline
\end{tabular}
\caption{The results of the $\chi^2$ fits to the first 3 and all modes but
$\omega$ in Table~\ref{tab:fit_simple}. The correlation between $|V_{ub}^L|$ and
$\epsilon_R$ in the two fits are 0.01 and 0.01.}
\label{tab:fit_Vubl_details_noS}
\end{table}

In Section~\ref{sec:newobs} we discuss the decay rate distributions. Besides
investigating the well known forward-backward asymmetry, we propose a
generalized two-dimensional asymmetry as a new observable that would be
interesting to measure. Additionally we explore the possibility to extract the
full information on the differential rate by considering asymmetries in all
three angles simultaneously.
In Section~\ref{sec:sumrule} we discuss the theoretical uncertainties
in existing form factor calculations. Using results from a light-cone sum rule
calculation~\cite{Ball:2004rg}, we estimate the correlations among the
uncertainties. Then we perform a simultaneous fit to a (simplified) series
expansion parametrization of the form factors. In Section~\ref{sec:observables}
we discuss the best theoretical predictions to extract information
on right-handed currents. We investigate the
discriminating contour for the two dimensional asymmetry.  We
estimate the sensitivity both with the current
$B$-factory data, as well as with the anticipated Belle~II
dataset to compare the various observables. 
We use this information in Section~\ref{sec:global_fit} to explore the impact of
the sensitivity to right-handed currents by performing global fits
simultaneously to $|V_{ub}^L|$ and $\epsilon_R$ assuming different scenarios for
both the current $B$-factory as well as expected Belle~II dataset.
Section~\ref{sec:conclusions} contains our conclusions.

\section{Possible observables}
\label{sec:newobs}

Starting from the Lagrangian in Eq.~(\ref{finalLNew}), the $B\to\rho\ell\bar\nu$
decay is described by replacing in the matrix element the vector ($V$) and the
three axial-vector ($A_{0,1,2}$) form factors via
\begin{equation}
V \to (1+\epsilon_R)\,V\,, \qquad  A_i \to (1-\epsilon_R)\,A_i\,.
\end{equation}
(If ${\rm Im}\, \epsilon_R=0$ then this can be done in the decay rate, too.)
Recently, the similar $B\to K^*\ell^+\ell^-$ decay has received a lot of
attention, in which case the decay distributions are in exact analogy with $B\to
\rho\ell\bar\nu$ (assuming that the neutrino is reconstructed).  It has been
advocated~\cite{Matias:2012xw} to use the form factor relations proposed in the
heavy quark limit~\cite{Charles:1998dr, Burdman:1998mk} to construct
observables, which are ratios of terms in the fully differential decay
distribution, to optimize sensitivity to new physics.  However, the size of
perturbative and nonperturbative corrections to these relations are subject to
discussions~\cite{Bauer:2002aj, Beneke:2002ph, Jager:2012uw}.  Thus, other
recent papers~\cite{Hambrock:2013zya} also have to resort to some extent to QCD
sum rule calculations to estimate the corrections to the form factor relations,
which we discuss in Sec.~\ref{sec:sumrule}.

\subsection{The general parameterization}

The fully differential decay rate for the four-body decay $B \to \rho (\to \pi
\pi) \ell^- \bar{\nu}_\ell$ can be written in terms of four variables. These are
conventionally chosen as the momentum transfer to the dilepton system, $q^2$,
and three angles describing the relative orientation of the final state
particles. As usual, we choose $\theta_V$ as the angle of the $\pi^+$ in the
$\rho$ restframe with respect to the $\rho$ direction in the $B$ restframe.
Similarly, $\theta_\ell$ is the angle of the $\ell^-$ in the dilepton
restframe with respect to the direction of the virtual $W^-$ in the $B$
restframe. Finally $\chi$ is the angle between the decay planes of the hadronic
and leptonic systems in the $B$ restframe. This convention coincides with the
usual definition in the similar flavor-changing neutral-current decay $B\to
K^*(\to K\pi) \ell^+ \ell^-$~\cite{Jager:2012uw,Descotes-Genon:2013vna}. The
fully differential rate is
\begin{align}\label{J19}
 &\frac{\text{d}\Gamma}{\text{d}q^2\, \text{d}\cos\theta_V\,
 \text{d}\cos\theta_\ell\, \text{d}\chi}
 = \frac{G_F^2 |V_{ub}^L|^2 m_B^3 }{2 \pi^4} \nonumber\\
&\times \bigg\{ J_{1s}  \sin^2 \theta_V + J_{1c} \cos^2 \theta_V \nonumber\\
&+ ( J_{2s} \sin^2 \theta_V + J_{2c} \cos^2 \theta_V) \cos 2\theta_\ell \nonumber\\
&+ J_{3} \sin^2 \theta_V \sin^2 \theta_\ell \, \cos 2 \chi\nonumber\\
&+ J_{4} \sin 2 \theta_V \sin 2  \theta_\ell \, \cos \chi
+ J_{5} \sin 2 \theta_V \sin \theta_\ell \, \cos \chi \nonumber\\
&+ ( J_{6s} \sin^2 \theta_V  + J_{6c} \cos^2 \theta_V ) \cos \theta_\ell \nonumber\\
&+ J_{7} \sin 2 \theta_V \sin \theta_\ell  \, \sin \chi 
+ J_{8} \sin 2 \theta_V \sin 2 \theta_\ell \, \sin \chi  \nonumber\\
&+ J_{9} \sin^2 \theta_V \sin^2 \theta_\ell \, \sin 2 \chi
\bigg\}\,.
\end{align}
Our convention for the ranges of the angular variables are $\chi \in[0,2\pi]$,
$\theta_\ell \in[0,\pi]$, $\theta_V \in[0,\pi]$. Switching $\chi \to \chi-\pi$,
so that $\chi \in [-\pi, \pi]$, customary in $B\rightarrow K^*\ell^+ \ell^-$,
amounts to a sign flip in the terms
\beq
\{J_4,\, J_{5},\, J_{7},\, J_8\} \to \{-J_4,\, -J_{5},\, -J_{7},\, -J_8\}\,.
\eeq
The dependence on $q^2$, as well as that on all form factors and on the NP
parameter $\epsilon_R$, is contained in the 12 dimensionless $J_i (q^2,
\epsilon_R)$ functions.  For the Lagrangian in Eq.~(\ref{finalLNew}), some
simplifications occur
\beq
    J_{1s} = 3 J_{2s}\,, \qquad 
    J_{1c} = - J_{2c}\,, \qquad J_7 = 0\,,
\eeq
and additionally $J_{6c} = 0$ for massless leptons. While the functions
$J_{7,8,9}$ are proportional to $\text{Im}\, \epsilon_R$, the other $J_i$
functions start with $(\text{Im}\, \epsilon_R)^2$ and $\text{Re}\, \epsilon_R$,
and so they are mainly sensitive to $\text{Re}\, \epsilon_R$.   Partially
integrated rates can be found in Appendix~\ref{sec:app_partial}.

An important difference between $B \to \rho \ell \bar\nu$ and $B \to K^*
\ell^+\ell^-$ is that in the former case the leptonic current is constrained to
be left-handed, and in the latter case several operators contribute already in
the SM, thus it is more compelling to study all possible NP contributions.
(Right-handed $\ell\bar\nu$ couplings are severely constrained, e.g., by Michel
parameter analyses.)  The rate corresponding to switching from left-handed to
right-handed leptonic current is obtained by the replacement $\theta_\ell \to
\theta_\ell - \pi$, resulting in a sign flip of the terms
\beq
\{J_5,\, J_{6s},\, J_{6c},\, J_7\} \to \{-J_5,\, -J_{6s},\, -J_{6c},\, -J_7\}\,.
\eeq
(As well as multiplication by the square of the right-handed coupling;
neglecting lepton masses, there is no interference between the two lepton
couplings.)
This difference can only be seen in an angular analysis, as it does not
contribute after integration over the angles. The $q^2$ spectrum depends on $2 J_{1s} + J_{1c} - (2
J_{2s} + J_{2c})/3$ and hence is insensitive to the chirality of the lepton
current.

In $B\to K^* \ell^+ \ell^-$ decay, a set of ``clean
observables" were proposed~\cite{Matias:2012xw}, which can be calculated
model independently in the SM, if the so-called ``non-factorizable"
contributions dominate the form factors~\cite{Bauer:2002aj}.
These observables are ratios of the $J_i$ functions, constructed so that these 
non-factorizable contributions cancel at each value of $q^2$, while there are
corrections from power suppressed effects as well as calculable ``factorizable"
contributions.  The cancellation of the non-factorizable contributions arises
because in the heavy $b$-quark limit, the number of independent nonperturbative
parameters is reduced due to the symmetries of SCET~\cite{Bauer:2000ew,
Bauer:2001yt}. However, even in this case, symmetry breaking corrections may be
a significant limitation in practice~\cite{Jager:2012uw}.
In the following we explore the possibilities of constructing observables
sensitive to a right-handed current. 

A fully differential analysis in four-dimensions, as required for the
determination of the $J_i$ in bins of $q^2$ for the calculation of the ``clean
observables" is experimentally challenging: an unbinned fit to the
four-dimensional decay rates requires parametrizing the background components
and their correlations adequately and when faced with this problem
experimentalists often choose alternative approaches, e.g., projections are
analyzed (see Refs.~\cite{Aubert:2007rs,Dungel:2010uk}) or event probabilities
are assigned (see, e.g., Ref.~\cite{CLEO:2011ab}). Both methods are complicated,
and as we are interested in the search for right-handed currents, corresponding
to constraining a single unknown parameter, we explore simpler variables, which
amount to counting experiments in different regions of phase space.

\subsection{One- and generalized two-dimensional asymmetries}

It is well known that the forward-backward asymmetry is sensitive to the chiral
structure of currents contributing to a decay,
\begin{equation}
A_{\rm FB} = \frac{\int_{-1}^0 \text{d}\cos\theta_\ell
  (\text{d}\Gamma/\text{d}\cos\theta_\ell)
  - \int_0^1 \text{d}\cos\theta_\ell
  (\text{d}\Gamma/\text{d}\cos\theta_\ell)}
  {\int_{-1}^1 \text{d}\cos\theta_\ell\, (\text{d}\Gamma/\text{d}\cos\theta_\ell)}\,.
\label{eq:afb}
\end{equation}
We study the sensitivity of this variable to $\epsilon_R$ in
Sec.~\ref{sec:observables}, after discussing the form factor inputs used.  The
one-dimensional distributions in $\chi$ and $\theta_V$ are symmetric, and hence
it is not possible to construct asymmetry-type observables with good sensitivity
to $\epsilon_R$ from these one-dimensional distributions.

Next, we integrate over one of the three angles, which reduces the number of
contributing $J_i$. We achieve the best sensitivity by integrating over the
angle $\chi$, which leaves us with
\begin{align}
 &\frac{\text{d}\Gamma}{\text{d}q^2\, \text{d}\cos\theta_V\,\text{d}\cos\theta_\ell}
 = \frac{G_F^2\, |V_{ub}^L|^2\, m_B^3 }{\pi^3}\, \bigg\{ J_{1s} \sin^2 \theta_V\nonumber\\
&+ J_{1c} \cos^2 \theta_V 
+( J_{2s} \sin^2 \theta_V + J_{2c} \cos^2 \theta_V)\cos 2\theta_\ell \nonumber\\
&+ ( J_{6s} \sin^2 \theta_V  + J_{6c} \cos^2 \theta_V ) \cos \theta_\ell 
\bigg\}\,\label{d2Gamma}
\end{align}
and $J_{6c} = 0$ for massless leptons.  This limits the possible observables
substantially, and none of the ``clean observables" sensitive to $\epsilon_R$
are accessible from measurement of this triple differential rate only.

To optimize the sensitivity from this class of measurements, we introduce new
observables,
\begin{equation}\label{NewVariable}
  S = \frac{A-B}{A+B} \,,
\end{equation}
where $A$ and $B$ are the decay rates in two regions in the
$\{\cos\theta_\ell,\, \cos\theta_V\}$ parameter space, chosen such that $S
\simeq 0$ in the SM. This is a generalization of the forward-backward asymmetry,
which may have increased sensitivity to $\epsilon_R$. To improve the statistical
precision, we integrate over a suitably chosen interval of $q^2$.  Given the
available constraints on the form factors, we integrate over $0 \leq q^2 \leq
12\, \text{GeV}^2$ to balance between experimental and theoretical
uncertainties.

\begin{table*}[tb]
  \setlength{\tabcolsep}{6pt}
\begin{tabular}{ccccc}
\hline\hline
$J_i$ & $\eta_i^\chi$ & $\eta_i^{\theta_\ell}$ & $\eta_i^{\theta_V}$ &
normalization $N_i$ \\
\hline
$J_{1s}$ & $\{+\}$ & $\{+,a,a,+\}$ & $\{-,c,c,-\}$ & $2\pi(1)2$ \\
$J_{1c}$ & $\{+\}$ & $\{+,a,a,+\}$ & $\{+,d,d,+\}$ & $2\pi(1)(2/5)$ \\
$J_{2s}$ & $\{+\}$ & $\{-,b,b,-\}$ & $\{-,c,c,-\}$ & $2\pi(-2/3) 2$ \\
$J_{2c}$ & $\{+\}$ & $\{-,b,b,-\}$ & $\{+,d,d,+\}$ & $2\pi(-2/3)(2/5)$ \\
$J_3$ & $\{+,-,-,+,+,-,-,+\}$ & $\{+\}$ & $\{+\}$ & $4 (4/3)^2$ \\
$J_4$ & $\{+,+,-,-,-,-,+,+\}$ &$\{+,+,-,-\}$ & $\{+,+,-,-\}$ & $4(4/3)^2$ \\
$J_5$ & $\{+,+,-,-,-,-,+,+\}$ & $\{+\}$ & $\{+,+,-,-\}$ & $4(\pi/2)(4/3)$ \\
$J_{6s}$ & $\{+\}$ & $\{+,+,-,-\}$ & $\{-,c,c,-\}$ & $2\pi(1)2$ \\
$J_{6c}$ & $\{+\}$ & $\{+,+,-,-\}$ & $\{+,d,d,+\}$ & $2\pi(1)(2/5)$ \\
$J_7$ & $\{+,+,+,+,-,-,-,-\}$ & $\{+\}$ & $\{+,+,-,-\}$ & $4(\pi/2)(4/3)$ \\
$J_8$ & $\{+,+,+,+,-,-,-,-\}$ & $\{+,+,-,-\}$ & $\{+,+,-,-\}$ & $4(4/3)^2$ \\
$J_9$ & $\{+,+,-,-,+,+,-,-\}$ & $\{+\}$ & $\{+\}$ & $4(4/3)^2$ \\
\hline\hline
\end{tabular}
\caption{Definition of the asymmetries in the three angles in bin-size of
$\pi/4$, see Eq-~\eqref{fact}. The $\pm$ signs denote $\pm1$, and $\{+\}$ denotes $+1$ in all
entries in a given column. Simple choices are $a=1-1/\sqrt2$, $b=a\sqrt2$,
$c=2\sqrt2-1$, and $d=1-4\sqrt2/5$.}
\label{tab:asym}
\end{table*}

It is important to estimate a reliable theoretical uncertainty for the asymmetry
$S$.  A priori, one may think that the hadronic uncertainties in the numerator
and the denominator cancel in the ratio to a large extent.  As it is
shown below, we cannot simply assume such a cancellation of nonperturbative
uncertainties in the ratios of the binned rates, as the considered $q^2$ region
is sizable. We develop a model for the uncertainties and correlations among
the binned rates, using available calculations of the form factors.
The optimal separation which discriminates between the two regions, $A$ and $B$,
depends on this choice of the $q^2$ range. Thus it is crucial to test the
sensitivity of the result to nonperturbative uncertainties.  

\subsection{\boldmath Binned measurements of the $J_i$ coefficients}

The previous approaches have the limitations of not allowing to chose the
numerator and denominator arbitrarily in terms of the $J_i$ functions. The
extraction of the full set of these coefficients is experimentally challenging,
and we propose a method that may allow for a better extraction of these
coefficients.  (The determination of a subset of the $J_i$ coefficients from
binned rates was explored in Ref.~\cite{DeCian}.)  We then investigate the
sensitivity of arbitrary ratios of the $J_i$. 

The form of the differential rate in Eq.~\eqref{J19} enables us to separate each
coefficient function $J_i$ from binning the three angles in fairly large,
$\pi/2$ size, bins.  Since some bin-boundaries need to be at half-integer
multiples of $\pi/2$, we use a notation where $\chi$ and $\theta_{V,\ell}$ are
split into 8 and 4 equal bins of size $\pi/4$, respectively.  We can then write
\begin{equation}
J_i = \frac1{N_i}\, \sum_{j=1}^8\, \sum_{k,l=1}^4
  \eta_{i,j}^\chi\, \eta_{i,k}^{\theta_\ell}\, \eta_{i,l}^{\theta_V}
  \Big[ \chi^{(j)} \otimes \theta_\ell^{(j)} \otimes \theta_V^{(k)} \Big]\,,
  \label{fact}
\end{equation}
where the $\eta^\alpha_{i,n}$ are weight factors listed in Table~\ref{tab:asym},
and the term in square brackets denotes the partial rate in the bin labeled by
$jkl$.  Thus, one can obtain the coefficient functions $J_i$ at each value of
$q^2$, or in bins of $q^2$, $\bar J_i = \int_{\Delta q^2} \text{d}q^2 J_i$.
Taking ratios of $J_i$-s to cancel some experimental and theoretical
uncertainties, as well as the dependence on $|V_{ub}|$, leads to observables
closely related to the $P_i$ ``clean observables" in the literature. 

Here a somewhat different way of extracting all the $J_i$ is proposed. For $B\to
K^* \ell^+ \ell^-$ the angular folding technique was used~\cite{Aaij:2013iag} to
extract these observables either via a counting  method or via a full unbinned
fit. An unbinned analysis is more difficult for $B \to \rho \ell \bar \nu$ due
to the sizable $B \to X_u \ell \bar \nu$ background. This background cannot be
assumed to be completely uncorrelated in the three-angle differential
distribution, which complicates the parametrization of the background
considerably.  Extracting all $J_i$-s from a binned asymmetry enables one to
perform cross-checks with the angular folding technique for all
observables~\cite{DeCian}, also in the case of $B\to K^*\ell^+ \ell^-$.

In analogy with Ref.~\cite{Descotes-Genon:2013vna}, we define
\begin{align}
\langle P_1 \rangle_\text{bin} &= \frac12 \frac{\int_{\Delta q^2} \text{d}q^2 J_3}{\int_{\Delta q^2} \text{d}q^2 J_{2s}}\,,
  \label{clean_sens1}\\
\langle P_4'\rangle_\text{bin} &= \frac{\int_{\Delta q^2} \text{d}q^2 J_4}{\sqrt{- \int_{\Delta q^2} \text{d}q^2 J_{2s}\,\,\int_{\Delta q^2} \text{d}q^2 J_{2c} }}
  \,,\label{clean_sens2}\\
\langle P_5' \rangle_\text{bin} &= \frac12 \frac{\int_{\Delta q^2} \text{d}q^2 J_5}{\sqrt{- \int_{\Delta q^2} \text{d}q^2 J_{2s}\,\,\int_{\Delta q^2} \text{d}q^2 J_{2c}}}
  \,, \label{clean_sens3}
\end{align}
which are the most sensitive to a possible right-handed
current (in terms of theoretical uncertainties), while the other ``clean
observables" either vanish or are less sensitive to a right-handed current.
Furthermore, we find that we get best sensitivity for simple
ratios, defined as
\begin{equation}
  \langle P_{i,j} \rangle_\text{bin} = 
  \frac{\int_{\Delta q^2} \text{d}q^2 J_i}{\int_{\Delta q^2} \text{d}q^2 J_j}
  \,.\label{clean_sens4}
\end{equation}
In particular, some coefficients which depend on all three angles have good
sensitivities, $\langle P_{3,4}\rangle$, $\langle P_{3,5}\rangle$, and $\langle
P_{5,4}\rangle$.

We can now constrain ${\rm Im}\, \epsilon_R$ as well.  The above defined
$\langle P_i \rangle$ observables only have quadratic dependence on $\text{Im}\,
\epsilon_R$, and for $\langle P_1 \rangle$, $\langle P_4' \rangle$, $\langle
P_{3,4} \rangle$ these contributions from the imaginary part start at order
$\text{Re}\, \epsilon_R$, and hence are strongly suppressed. However, from the
linear dependence in $J_{7,8,9}$ we can construct sensitive observables to ${\rm
Im}\, \epsilon_R$, with a quadratic dependence on the real part, namely $
\langle P_{8,5}\rangle$ and $\langle P_{9,5}\rangle$. Furthermore it is
interesting to look at $\langle P_{8,3}\rangle $, which starts with a linear
dependence on the real part, but has a very large slope with respect to
$\epsilon_R$, while at the same time a very small theoretical uncertainty.

The next section discusses the light-cone sum rule calculation, the correlations
among the form factors.  Then we derive the optimal two-dimensional asymmetry,
$S$, and subsequently return to the sensitivities in $\epsilon_R$ obtainable
through all observables discussed.

\section{Form Factor calculation and fit}
\label{sec:sumrule}

\subsection{The series expansion (SE) and the
simplified series expansion (SSE)}

It has long been known that unitarity and analyticity impose strong constraints
on heavy meson decay form factors~\cite{Grinstein:1992hq, Boyd:1994tt,
Arnesen:2005ez, Becher:2005bg, Bourrely:2008za}.  We use a series expansion,
also known as the $z$ expansion, to describe the form factor shape over the full
range of the dilepton invariant mass.  Using this expansion for a vector meson
in the final state, instead of a pseudoscalar, requires additional
assumptions~\cite{Bharucha:2010im}, and we investigate the corresponding
uncertainties.  In this paper we expand the form factors directly, instead of
the helicity amplitudes.

The series expansion uses unitarity to constrain the shape of the form factors,
and implies a simple and well-motivated analytic parametrization over the full
range of $q^2$.  The form factors are written as
\begin{align}\label{VAexp}
V(q^2) &=  \frac{1}{B_V(q^2)\, \Phi_V(q^2)} \,
  \sum_{k=0}^K \alpha^V_k\, z(q^2,\, q^2_0)^k \,, \nn\\*
A_i(q^2) &= \frac{1}{B_{A_i}(q^2)\, \Phi_{A_i}(q^2)} \,
  \sum_{k=0}^K \alpha^{A_i}_k\, z(q^2,\, q^2_0)^k \,,
\end{align}
where unitarity constrains the shapes of the form factors by predicting
$\Phi_{F}(q^2)$, $F = \{V,\, A_i\}$, and also bounds the coefficients of the
expansion in powers of the small parameter, $z(q^2, q^2_0)$, schematically as $\sum_{k=0}^\infty \big(\alpha_k^F\big)^2 < 1$. (For $q^2$ relevant for semileptonic $B$ decay, $|z(q^2, q^2_0)|<1$.)
In Eq.~(\ref{VAexp}) the variable
\begin{equation}
z(q^2,q^2_0) = \frac{\sqrt{q^2_+-q^2}-\sqrt{q^2_+-q^2_0}}
  {\sqrt{q^2_+-q^2}+\sqrt{q^2_+-q^2_0}}\,,
\end{equation}
maps the real $q^2$ axis onto the unit circle, 
$q_0^2$ is a free parameter, and $q^2_\pm \equiv (m_B\pm m_\rho)^2$. The range
$-\infty < q^2 < q^2_+$ is mapped onto the $-1 < z(q^2<q^2_+,\, q^2_0) < 1$ line
segment on the real axis inside the unit disk, while the branch cut region
corresponding to $B\rho$ pair creation, $q^2 > q_+^2$, maps onto the unit
circle, $|z(q^2>q^2_+,\, q^2_0) | = 1$.  The $q_0^2$ parameter of this
transformation is usually chosen as
\begin{equation}
  q^2_0 = (m_B + m_\rho)\, (\sqrt{m_B} - \sqrt{m_\rho} )^2,
\end{equation}
so that for the physical $q^2$ range of $B\to \rho\ell\bar\nu$ decay, $0 \leq
q^2\leq q^2_-$, the expansion parameter is minimal, $|z(q^2,q^2_0)| < 
\big(1-\sqrt[4]{1-q_-^2/q_+^2}\big) \big/ \big(1+\sqrt[4]{1-q_-^2/q_+^2}\big) 
\approx 0.1$.
The so-called Blaschke factors in Eq.~(\ref{VAexp}) for each form factor are
\begin{equation}
B_F (q^2) \equiv \prod_{R_F} z(q^2,\, m_{R_F}^2)\,, \label{Blaschke}
\end{equation}
where $R_F$ are the sub-threshold resonances ($q_-^2 < m_{R_F}^2 < q_+^2$)
with the quantum numbers appropriate for each form factor.  By construction,
$B_F(m_{R_F}^2) = 0$ and $|B_F(q^2)| = 1$ for $q^2>q_+^2$. The main shape
information is given by the functions~\cite{Bharucha:2010im}
\begin{align}
\Phi_F(q^2) &= \sqrt{\frac1{32 \pi \chi_F(n)}}\,
   \frac{q^2-q^2_+}{(q^2_+-q^2_0)^{1/4}}
   \left[\frac{z(q^2,0)}{-q^2}\right]^{(n+3)/2}  \nonumber\\
&\quad \times\left[\frac{z(q^2,q^2_0)}{q^2_0 - q^2}\right]^{-1/2}
    \left[\frac{z(q^2,q^2_-)}{q^2_--q^2}\right]^{-3/4} . \label{phifunc}
\end{align}
The only form factor dependent quantity is $\chi_F(n)$, which is related to the
polarization tensor $\Pi_{\mu\nu}(q^2)$ at $q^2 = 0$, and $n$ is the number of
derivatives (subtractions) necessary to render the dispersion relation finite.
This function is calculable in an operator product expansion.  Since it is an
overall constant which does not affect the shapes of the form factors (and we do
not use a constraint on $\sum\alpha_i^2$), we can absorb this quantity into the
fit parameters $\alpha_i$. In contrast, the number of required subtractions $n$
influences the shape information. For the longitudinal part, involving
$A_0$, one subtraction is necessary, while for the transverse part of the
vector and axialvector current, involving the form factors $A_1$,
$A_2$, and $V$, two subtractions are needed~\cite{Bharucha:2010im}.

While constraining the shapes of the form factors, several uncertainties need to
be considered. Using analyticity requires the form factor to be free of poles
and branch cuts in the region $q^2_- < q^2 < q^2_+$, which is not true in
reality.  In the analysis of each form factor, $F$, resonances $R_F$ with
appropriate quantum numbers appear as sub-threshold singularities.  Their
effects can be eliminated by dividing with the Blaschke factors in
Eq.~\eqref{Blaschke}. However, some of the resonance are fairly broad, and their
masses, $m_{R_F}$, have uncertainties.  We checked that the final result is not
too sensitive to variations of the resonance masses by $\pm 100$\,MeV.  

Besides resonances, there are also branch cuts in the range $q^2_- < q^2<q_+^2$,
corresponding to multi-body states, such as $B + n \pi$ below the $B + \rho$
threshold.  This does not occur for $B\to \pi\ell\bar\nu$, and they cannot be
eliminated as easily as the poles.  Using a model for the branch
cuts~\cite{Caprini:1995wq}, we estimate in Appendix~\ref{bc_uncertainty} how the
unitarity bound changes numerically. We find that the expansion parameters,
$\alpha_i^F$, which would distort the shapes of the form factors, change at most
at a few percent. As these coefficients multiply small numbers, $|z(q^2,q_0^2)|
\lesssim 0.1$, we find that neglecting branch cuts does not change the form
factor shape significantly, as it probably mainly affects the saturation of
higher order terms in the expansion.

Another potential complication is due to the $\rho$ meson's substantial width,
which allows nonresonant $B\to \pi\pi \ell \bar\nu$ decay to contribute to the
$B\to \rho\ell\bar\nu$ signal. This can be handled using standard experimental
techniques, and a measurement of $B\to \pi^0\pi^0 \ell \bar\nu$ can be used to
constrain this background, as $\rho^0 \to \pi^0 \pi^0$ is forbidden.
Narrower cuts on the $\rho$ mass window can also reduce this uncertainty,
especially using larger data sets in the future.  

Throughout this paper we refer to the approach described so far as the series
expansion (SE).  We do not recalculate the bound on the expansion coefficients,
which shows that the expansion to linear order is a good
approximation~\cite{Bharucha:2010im}. However, we perform the fit both to linear
and to quadratic order and investigate from this the convergence behavior of the
SE.  As a cross-check of possible shape information bias with respect to
the input data, we also use the proposed simplified series expansion
(SSE)~\cite{Bourrely:2008za, Bharucha:2010im}, which further tests uncertainties
related to the form factor shape.  It is obtained via the replacements
\begin{align}
  \Phi(q^2) &\to 1\,, \nn\\
  B(q^2) &\to P(q^2) = \frac{1}{1-q^2/m_R^2}\,. 
\end{align}
Since using the helicity basis for the form factors is theoretically
favored~\cite{Bharucha:2010im,Jager:2012uw}, we compare our SE and SSE
parametrizations in the form factor basis and fitting the helicity
basis with Ref.~\cite{Bharucha:2010im}.  We find consistent results with
all parametrizations.  Since all studied approaches are very compatible, we limit
ourselves to show results using the linear series expansion parametrization.

\subsection{Correlation assumptions for the form factors}

Ideally, any determination of the form factors should also provide their
correlations, in addition to the central values and uncertainties, as it is
crucial for predicting uncertainties of observables dependent on several form
factors. Unfortunately this is currently not available from either lattice QCD
or model calculations. We estimate these correlations in the light-cone QCD sum
rule (LCSR) results~\cite{Ball:2004rg, Hambrock:2013zya}. We distinguish two
different kinds of correlations, (i) correlations among the different form
factors at the same value of $q^2$; and (ii) correlations between different
values of $q^2$, for the same form factors.  In general larger correlation between the form
factors will result in larger correlations of the fit parameters, and hence more
precise predictions, while larger correlations for different values of $q^2$
lead to less precise predictions.

In Ref.~\cite{Ball:2004rg}, the uncertainties at $q^2 = 0$ are grouped into four
sources, presumed uncorrelated: $\Delta_{7P}$, $\Delta_{m_b}$, $\Delta_{L}$, and
$\Delta_{T}$. The values evaluated for $q^2=0$ are listed in
Table~\ref{tab:sumrule_uncert}, and are used in the following as an estimate of
the uncertainties over a larger range of $q^2$. We investigate the individual
contributions to these uncertainties and estimate the correlation among the form
factors.

\begin{enumerate}

\item The leading contributions for $\Delta_{7P}$ are from the uncertainties in
the distribution amplitude of the $B$ meson and their expansion in Gegenbauer
moments. The source of these two uncertainties is assumed fully correlated among
all $A_i$ and $V$ in the following. Wen can assess the contribution from these
sources to $\Delta_{7P}$ from Fig.~4 in Ref.~\cite{Ball:2004rg}. This helps to
estimate the amount of correlation stemming from this source.

\item External inputs, e.g., the values of $m_b$ and of the condensates are
fully correlated among the form factors. Ref.~\cite{Hambrock:2013zya} argued
that the duality parameter and Borel parameter should also be treated as
strongly correlated.

\item The uncertainties due to the vector and tensor decay constants of the
$\rho$ are also correlated among all form factors. They enter the same
correlation function, see Eqs.~(32)--(37) in  Ref.~\cite{Ball:2004rg}.

\end{enumerate}

\begin{table}[b]
\begin{tabular}{c|cccccc}
\hline\hline
Form factor, $F$ & $F(q^2=0)$  & $\Delta_{7P}$ & $\Delta_{m_b}$ & $\Delta_{L}$ & $\Delta_{T}$   \\
\hline
$V(0)$ & 0.323 & 0.025 & 0.007 & 0.005 & 0.013  \\
$A_0(0)$ & 0.303 & 0.026 & 0.004 & 0.009 & 0.006  \\
$A_1(0)$ & 0.242 & 0.020 & 0.007 & 0.004 & 0.010  \\
$A_2(0)$ & 0.221 & 0.018 & 0.008 & 0.002 & 0.011  \\
\hline\hline
\end{tabular}
\caption{The uncertainties from $\Delta_{7P}$, $\Delta_{m_b}$, $\Delta_{L}$, and
$\Delta_{T}$ from Ref.~\cite{Ball:2004rg}.}
\label{tab:sumrule_uncert}
\end{table}

From these considerations, we can assess the correlated uncertainties in each
contribution. In the following  a model is tested to predict the correlations
between the form factors. For this model, according to the list above, the
correlations between the $A_i$, and between the $A_i$ and $V$ are assumed to be
$\big\{ \rho_{7P}^{A_i},\, \rho_{m_b}^{A_i},\, \rho_L^{A_i},\, \rho_T^{A_i}
\big\} = \{ 0.6,\, 1.0,\, 1.0,\, 1.0 \}$ and $\big\{ \rho_{7P}^{V,A_i},\,
\rho_{m_b}^{V,A_i},\, \rho_L^{V,A_i},\, \rho_T^{V,A_i} \big\} = \{ 0.6, 1.0,\,
1.0,\, 1.0\, \}$. A full calculation of the form factors and the complete
determination of the correlations is beyond the scope of this paper. Hence our
estimate relies on the results given in that paper, and on our assumptions
listed above. A new determination of these input values in a separate analysis
would be useful.

The total covariance can in turn be written as
\begin{equation}
 C = C_{7P} + C_{m_b} + C_{L} + C_{T}\,,
\end{equation}
where $C_j$ is a $4 \times 4$ matrix of the form
\beq\arraycolsep 3pt
{\small \!\left(\! \begin{array}{cccc}
\big( \Delta_j^{V} \big)^2 &  \rho_j^{V,A_i}  \Delta_j^{V}  \Delta_j^{A_0} 
  & \rho_j^{V,A_i} \Delta_j^{V}  \Delta_j^{A_1} &  \rho_j^{V,A_i} \Delta_j^{V} \Delta_j^{A_2} \\[4pt]
\rho_j^{V,A_i} \Delta_j^{V}  \Delta_j^{A_0} &\big( \Delta_j^{A_{0}} \big)^2 
  & \rho_j^{A_i} \Delta_j^{A_{0}} \Delta_j^{A_{1}} & \rho_j^{A_i} \Delta_j^{A_{0}} \Delta_j^{A_{2}} \\[4pt]
\rho_j^{V,A_i} \Delta_j^{V}  \Delta_j^{A_1} &\rho_j^{A_i}  \Delta_j^{A_{0}} \Delta_j^{A_{1}} &\big( \Delta_j^{A_{1}} \big)^2 &\rho_j^{A_i}  \Delta_j^{A_{1}} \Delta_j^{A_{2}} \\[4pt]
\rho_j^{V,A_i} \Delta_j^{V}  \Delta_j^{A_2} & \rho_j^{A_i} \Delta_j^{A_{0}}
\Delta_j^{A_{2}} & \rho_j^{A_i}  \Delta_j^{A_{1}}  \Delta_j^{A_{2}}&\big(
\Delta_j^{A_{2}} \big)^2 \end{array} \!\right) \!.
}\eeq
This results in the correlation matrix for $\{V,A_0, A_1, A_2\}$ given by
\beq\arraycolsep 6pt
C=  \left( \begin{array}{cccc} 
1. & 0.65 & 0.71 & 0.72 \\
0.65 & 1. & 0.64 & 0.62 \\
0.71 & 0.64 & 1. & 0.72 \\
0.72 & 0.62 & 0.72 & 1. \end{array} \right) , 
\eeq
This estimate is derived at $q^2=0$, and we use it for $q^2 > 0$ as well.
Because of the constraints on the shapes of the form factors, no large change is
expected far from maximal $q^2$.  

The form factors at different values of $q^2$ are obtained from the same sum
rule, however, the various contributions are weighted differently by $q^2$; see
Eqs.~(32)--(37) in Ref.~\cite{Ball:2004rg}.  For values of $q^2$ farther from
one another, the correlation should decrease.  We implemented the leading order
formulae~\cite{Ball:2004rg}, which are consistent with the full results for the
shapes of the form factors, and the magnitude is also consistent within the
uncertainty of the full result.  We found that the correlation for different
values of $q^2$ only mildly depends on the separation, which we use below. 
Thus, uncertainties of a given form factor, $A_i$ or $V$, for different $q^2$
are estimated to be 80\% correlated, which is a bit more conservative then the
75\% correlation used in Ref.~\cite{Bharucha:2010im} (with a binning of $3
\text{ GeV}^2$, whereas we use $1 \text{ GeV}^2$ in our analysis).

\begin{figure*}[t]
\centerline{\includegraphics[width=.9\columnwidth]{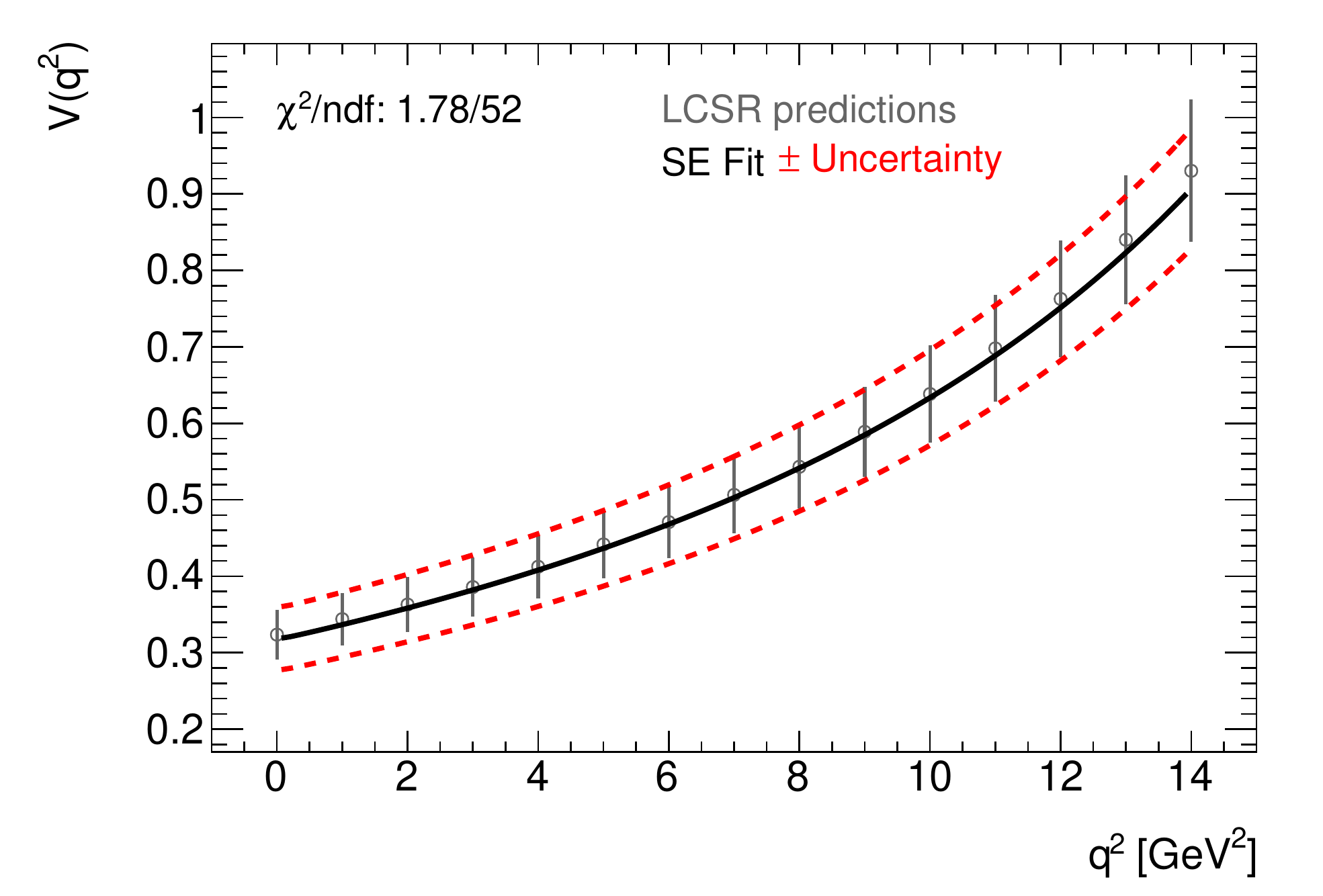}
\hfil \includegraphics[width=.9\columnwidth]{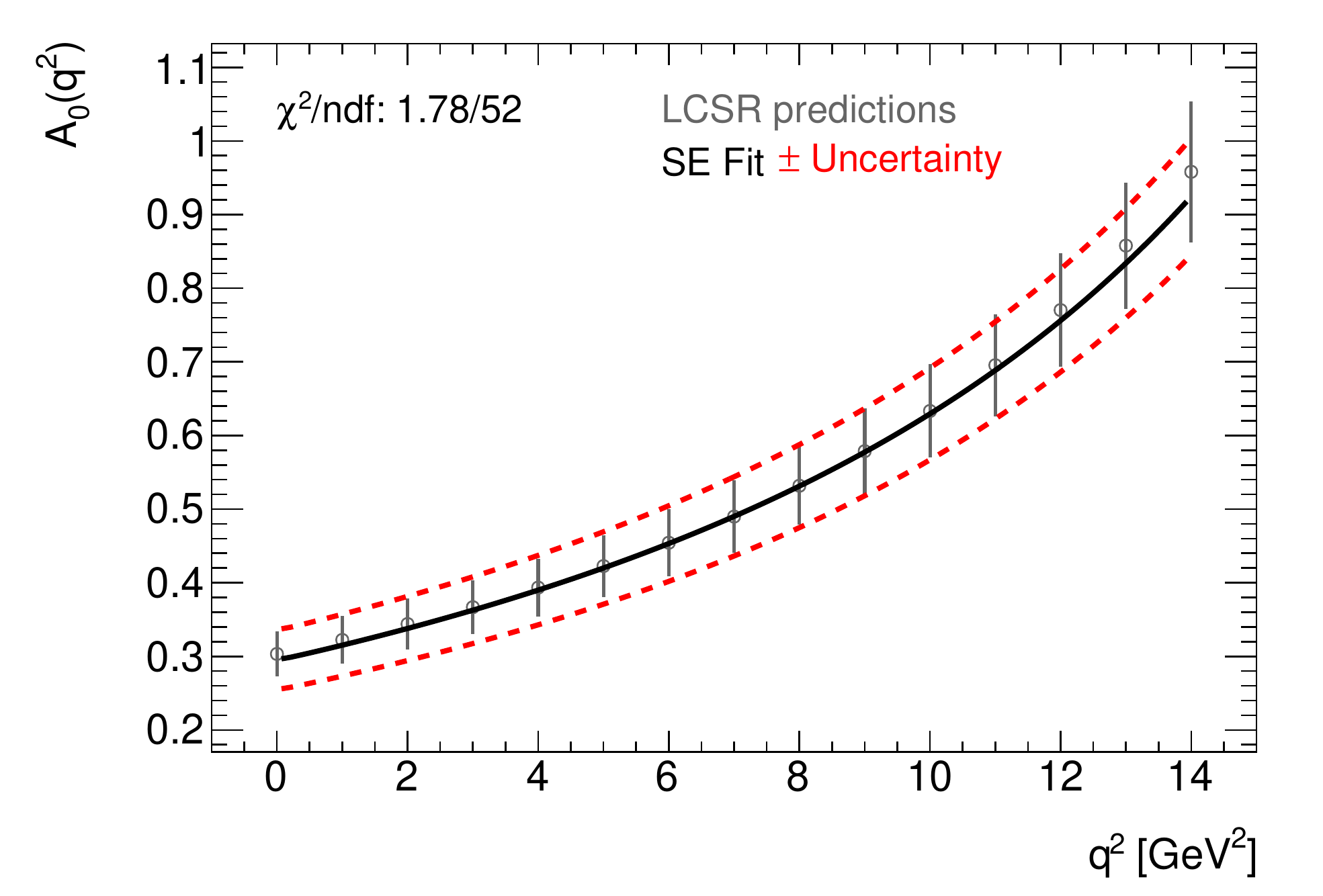}}
\centerline{\includegraphics[width=.9\columnwidth]{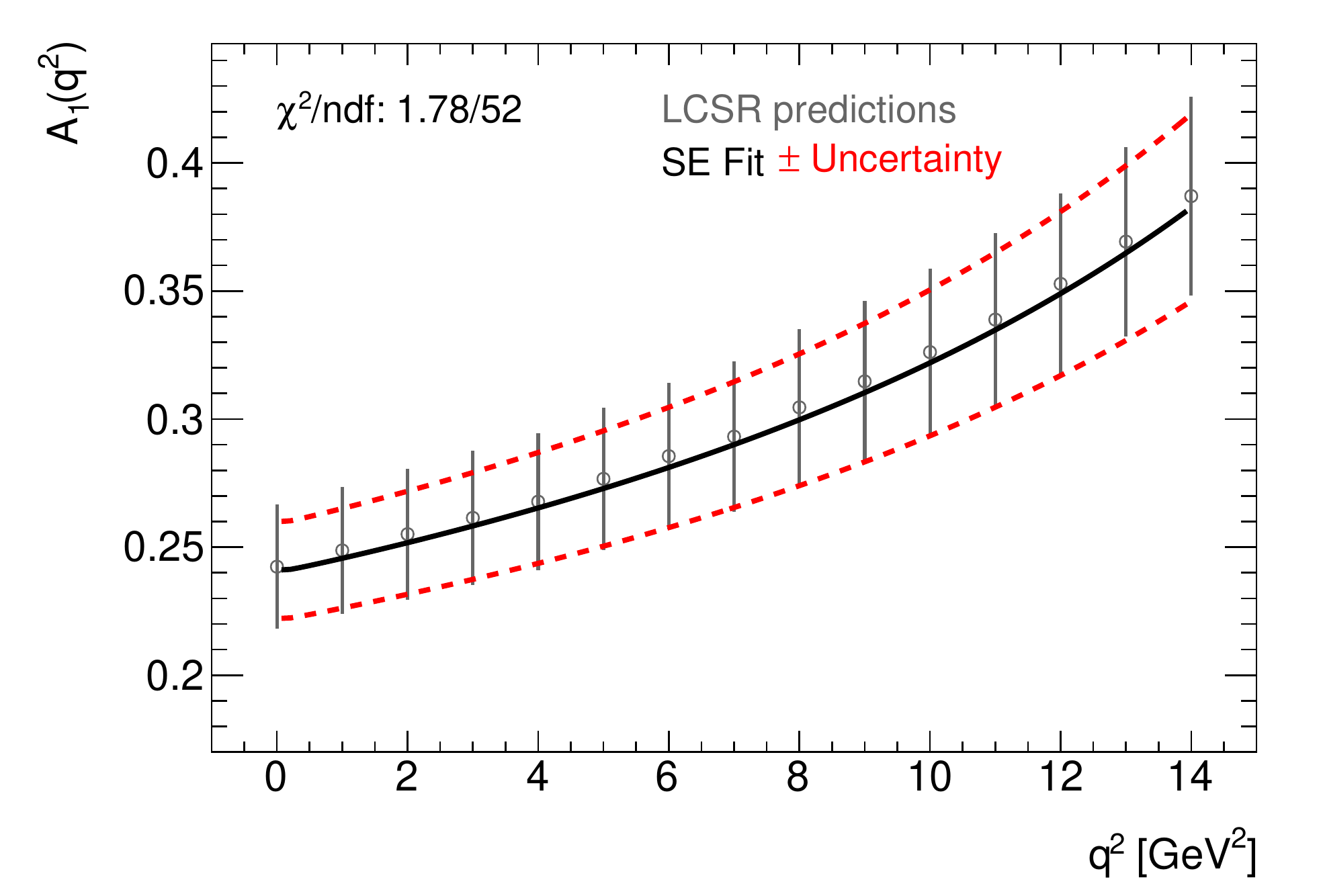}
\hfil \includegraphics[width=.9\columnwidth]{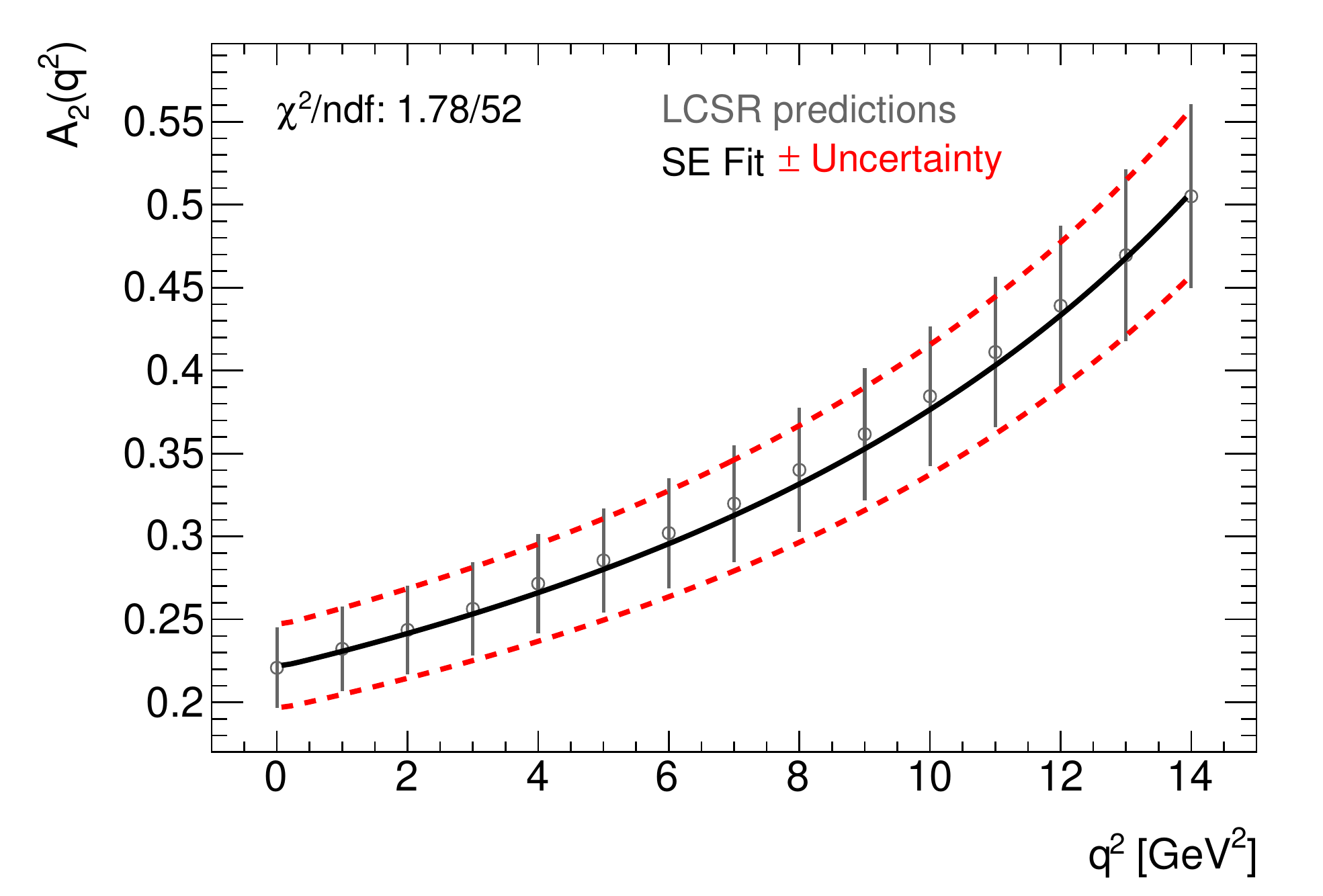}}
\caption{Simultaneous fits to the sum rule prediction of Ref.~\cite{Ball:2004rg}
using the linear full series expansion for the form factors: $V(q^2)$ (top
left), $A_0(q^2)$ (top right), $A_1(q^2)$ (bottom left), $A_2(q^2)$ (bottom
right). The solid black lines show the fitted form factor, the gray data points
show the fitted sum rule points, and the red dashed line shows the determined
uncertainty.}
\label{fig:SE_linear}
\end{figure*}

\subsection{\boldmath The $\chi^2$ fit for the SE and SSE parameters}

A simultaneous $\chi^2$ fit to all sum rule points of Ref.~\cite{Ball:2004rg}
assuming the correlations discussed in the previous section is performed. The
form factors are parametrized both in the full or in the simplified series
expansion, to either linear or quadratic order in $z$, for both the form factor
and the helicity amplitude basis. All of them show consistent results, and thus
we restrict ourselves to the SE at linear order. 
In \cite{Bharucha:2010im} a similar analysis with a less elaborate correlation 
treatment was performed. We find reasonable agreement with their form factor fit results, 
but due to the different correlation structure the uncertainties on physical observables differs
from this work.
The result of the fit to the linear SE is shown in
Fig.~\ref{fig:SE_linear}. The central values and uncertainties of the fit were
verified using ensembles of pseudo-experiments. (Varying the input assumptions
leaves the central values and uncertainties mostly stable, while the resulting
correlation matrix is slightly changed as one would expect.) The fitted values
for the full series expansion to linear order are listed in
Table~\ref{tab:sumrule_fitresults_lin}. The corresponding fit parameter
correlations are listed in Table~\ref{tab:sumrule_fitresults_lin2}.

\begin{table}[b]
\begin{tabular}{c|ccc|ccc}
\hline\hline
$F$ & $a_0^{F}$  & $a_1^{F}$  \\
\hline
$A_0$ & $-0.351 \pm 0.032$ &$1.250 \pm 0.147$  \\
$A_1$ & $-0.111 \pm 0.010$ &$-0.208 \pm 0.042$  \\
$A_2$ & $-0.138 \pm 0.014$ &$0.170 \pm 0.049$ \\
$V$	&$-0.366 \pm 0.034$ &$1.148 \pm 0.145$  \\
\hline\hline
\end{tabular}
\caption{Fit result for linear order SE.}
\label{tab:sumrule_fitresults_lin}
\end{table}

\begin{table}[b!]
\tabcolsep 4pt
\begin{tabular}{c|cccccccc}
\hline\hline
 & $a_0^V$ & $a_1^V$ & $a_0^{A_0}$ & $a_1^{A_0}$ & $a_0^{A_1}$ & $a_1^{A_1}$ & $a_0^{A_2}$ & $a_1^{A_2}$ \\
\hline
 $a_0^{A_0}$ & 1.00 & -0.86 & 0.77 & 0.35 & 0.74 & -0.26 & 0.78 & -0.57 \\
 $a_1^{A_0}$ & -0.86 & 1.00 & -0.60 & -0.27 & -0.58 & 0.20 & -0.61 & 0.44 \\
 $a_0^{A_1}$ & 0.77 & -0.60 & 1.00 & 0.31 & 0.86 & -0.31 & 0.85 & -0.62 \\
 $a_1^{A_1}$ & 0.35 & -0.27 & 0.31 & 1.00 & 0.39 & -0.14 & 0.39 & -0.28 \\
 $a_0^{A_2}$ & 0.74 & -0.58 & 0.86 & 0.39 & 1.00 & -0.49 & 0.86 & -0.63 \\
 $a_1^{A_2}$ & -0.26 & 0.20 & -0.31 & -0.14 & -0.49 & 1.00 & -0.31 & 0.22 \\
 $a_0^V$     & 0.78 & -0.61 & 0.85 & 0.39 & 0.86 & -0.31 & 1.00 & -0.82 \\
 $a_1^V$     & -0.57 & 0.44 & -0.62 & -0.28 & -0.63 & 0.22 & -0.82 & 1.00 \\
\hline\hline
\end{tabular}
\caption{Correlations for linear order SE.}
\label{tab:sumrule_fitresults_lin2}
\end{table}

The LCSR result is valid only for small $q^2$.  However, the $q^2$ distribution
changes by less than 1\% when fitted in the region $q^2 < 7\,\text{GeV}^2$ or
$q^2 < 14\,\text{GeV}^2$~\cite{Ball:2004rg}.  Since the measurements in
Ref.~\cite{Sibidanov:2013rkk} are in 4\,GeV bins, we restrict ourselves to
fitting the data in the range $q^2 < 12 \,\text{GeV}^2$ to optimize statistical
sensitivity while maintaining theoretical validity.

Our fitting procedure can perform a fit to several data sets.  In the future, a
combined fit to LCSR data, most reliable at low $q^2$, and lattice QCD data,
most reliable at high $q^2$, is desirable. That would constrain the shape of the
spectrum in an optimal way, and it would also test the compatibility of the two
approaches.  Since no reliable and precise lattice QCD calculation of $B\to
\rho$ form factors is available, this is left for future work.  The framework
developed in this work is capable to incorporating such future inputs, which
will also allow the whole experimental data set to be used without any
restriction on $q^2$.  Our fitting program is not restricted to $B\rightarrow
\rho$ form factors, but can easily be adopted for other processes using the
parametrizations discussed.

\begin{figure*}[t]
\centerline{
\includegraphics[width=.68\columnwidth, clip, bb=40 0 445 625]{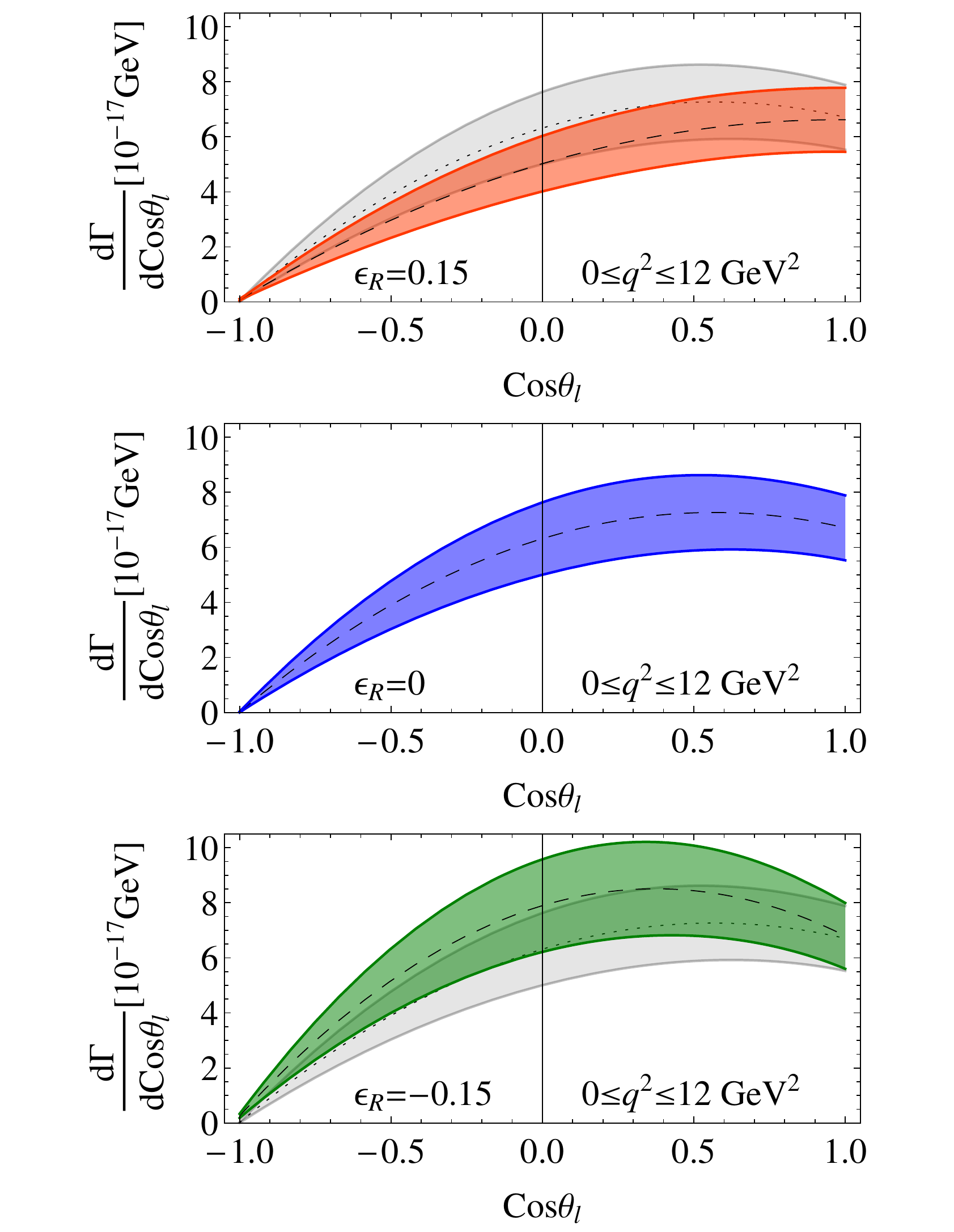}\hfill
\includegraphics[width=.68\columnwidth, clip, bb=40 0 445 625]{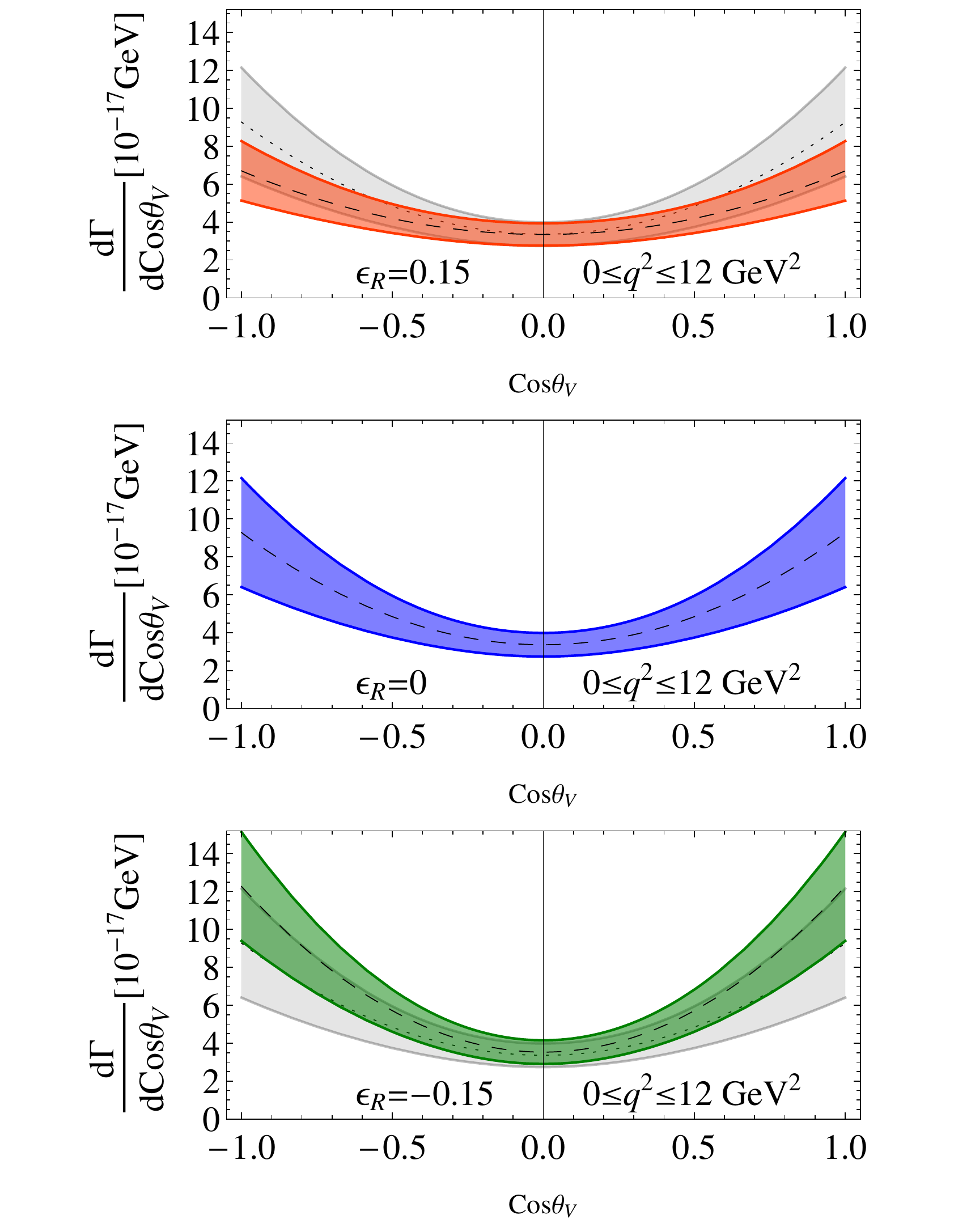}\hfill
\includegraphics[width=.68\columnwidth, clip, bb=40 0 445 625]{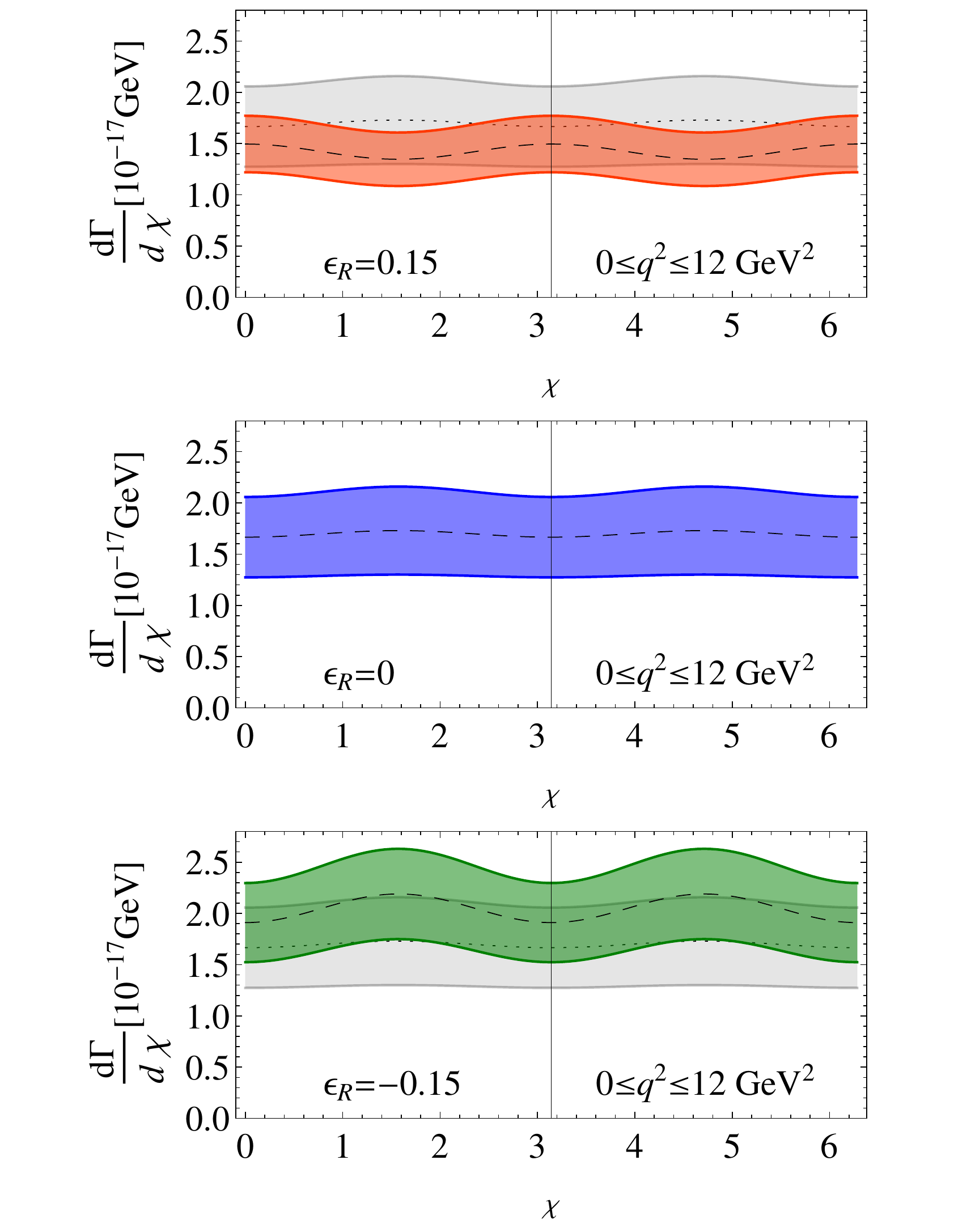}}  
\caption{The differential decay rate as a function of the helicity angles $\cos
\theta_{\ell}$ (left), $\cos \theta_{V}$ (middle), and $\chi$ (right), for
$\epsilon_R = 0$ (SM, middle line) and $\epsilon_R = \pm 0.15$.  The dashed
curves show the full series expansion to linear order. The shaded areas
correspond to the estimated theoretical uncertainty from the fit to the sum
rule prediction, taking into account the full correlation of the expansion
coefficients.}
\label{fig:SE_linear_coswl_cosv}
\end{figure*}

\section{Predictions of the observables}
\label{sec:observables}

In the following the theoretical predictions using the form factor input and
uncertainties from the last section are discussed.  The one-dimensional angular
distributions including the theoretical uncertainties are displayed in
Fig.~\ref{fig:SE_linear_coswl_cosv} with $|V_{ub}| = 4.2\times 10^{-3}$. The
large theoretical uncertainties due to the $B \to \rho$ form factor show the
necessity of constructing non-trivial observables to gain sensitivity for
right-handed contributions. 

The achievable sensitivity of the observables is estimated for 1\,ab$^{-1}$ and
50\,ab$^{-1}$ of integrated luminosity, corresponding to the available \babar
and Belle data sets and the anticipated Belle~II data. The experimental
sensitivities were estimated using the uncertainties of
Ref.~\cite{Sibidanov:2013rkk}, assuming that systematic uncertainties in
disjoint regions of phase space (e.g. between different bins of $J_i$) are fully
correlated. For 50\,ab$^{-1}$ an improvement of the systematic uncertainties of
a factor of $3$ is assumed, motivated by the improvements for $B \to X_u \ell
\bar \nu$ from Ref.~\cite{Aushev:2010bq} which face similar experimental
challenges.  The statistical uncertainties  were scaled to correspond to
1\,ab$^{-1}$ or 50\,ab$^{-1}$ integrated luminosity.  The expected sensitivity
for $\epsilon_R$ for each observable is characterized as a 68\% confidence
interval by using the Neyman construction assuming normal distributed
uncertainties. In practice, every experiment will have to derive these curves
from an ensemble of pseudo-experiments or asymptotic formulae with the specific
values of $\epsilon_R$ and proper experimental uncertainties incorporated. The
sensitivity  to a possible right-handed admixture is assessed  by the
interception of the uncertainty bands with the predicted SM value.  Experimental
and theoretical uncertainties are assumed to be independent, and addition in
quadrature is used to combine them.

\subsection{\boldmath Forward-backward asymmetry and the two-dimensional
asymmetry, $S$}

\begin{figure*}[t]
\centerline{
\includegraphics[width=.9\columnwidth]{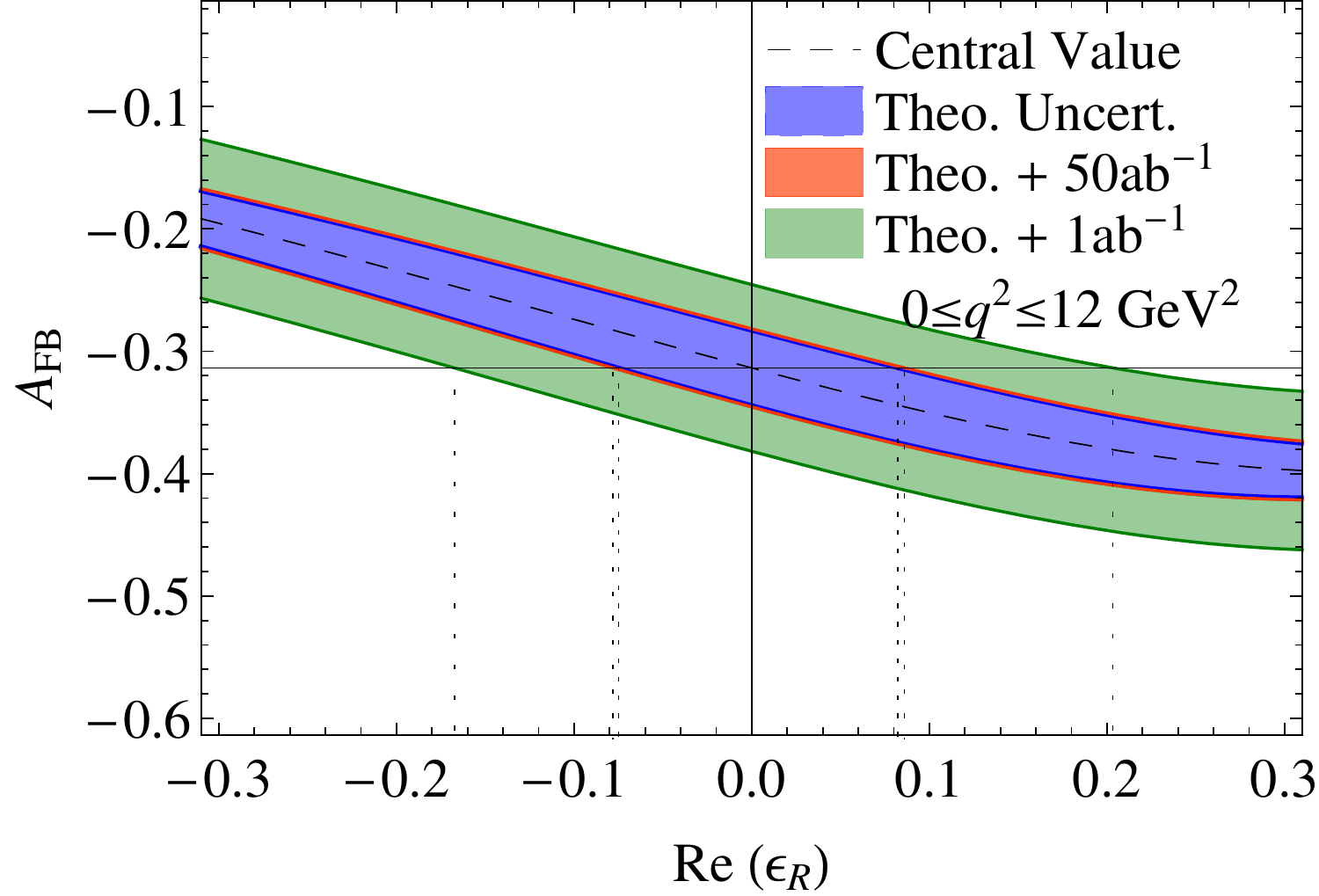}\hfil
\includegraphics[width=.9\columnwidth]{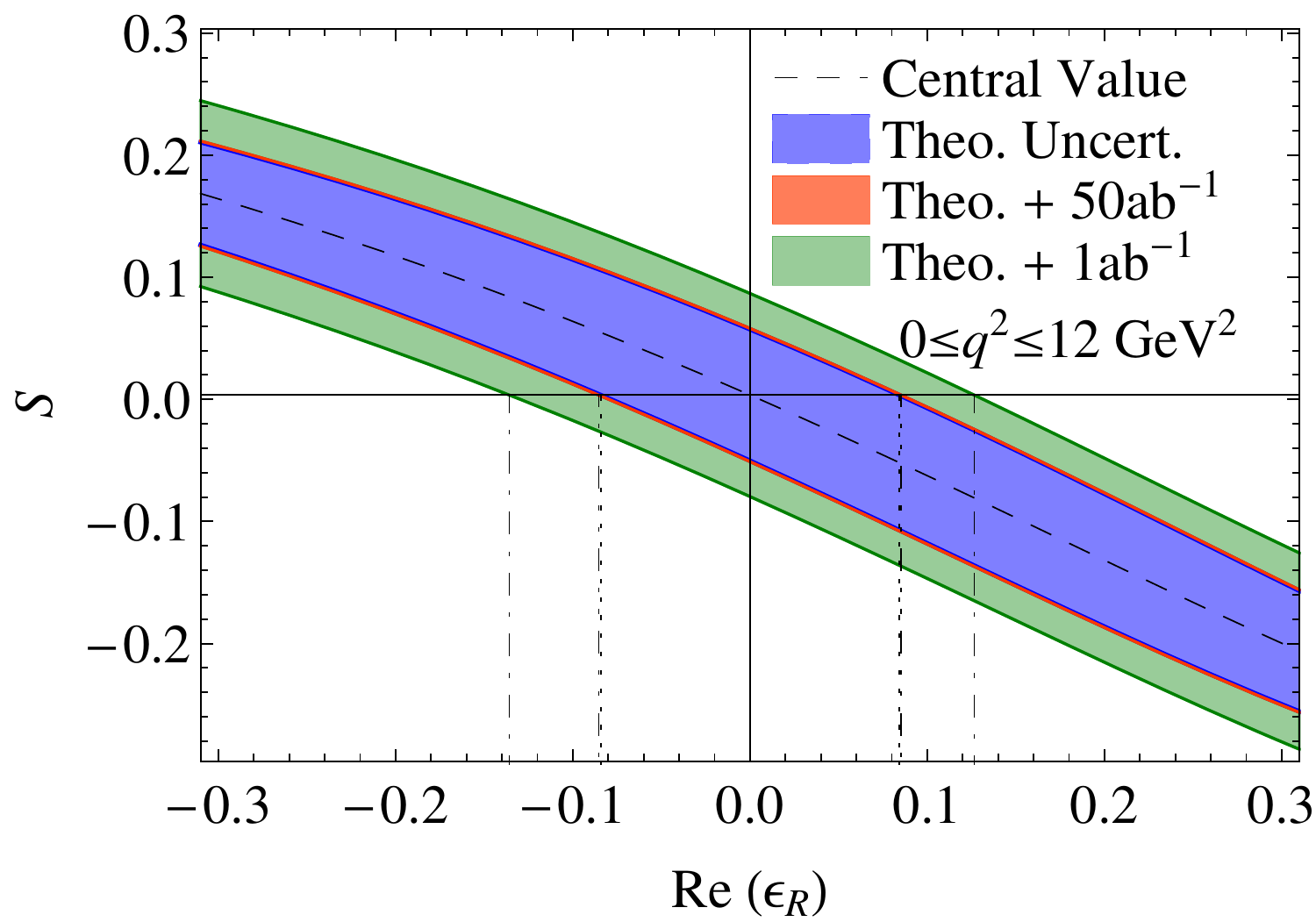}}
\caption{Predictions for the forward-backward asymmetry (left) and $S$ (right),
including theoretical uncertainties (blue band), and theory and experimental
uncertainties combined in quadrature for 50\,ab$^{-1}$ (orange) and 1\,ab$^{-1}$
(green).}
\label{fig:SE_lin_AFB_S}
\end{figure*}

Determining $A_{\rm FB}$ requires the measurements of the decay angle
$\theta_{\ell}$ and the predictions including uncertainty estimates are shown in
the left panel of Fig.~\ref{fig:SE_lin_AFB_S}. The central value is indicated by
dotted lines and the blue band shows the theory uncertainty, as derived in the
previous section. The red and green band show the total uncertainties for
$1\,{\rm ab}^{-1}$ and $50\,{\rm ab}^{-1}$ of integrated luminosity, and the
dashed vertical lines show the expected sensitivities assuming the SM. The
theoretical and experimental uncertainties for $1\,{\rm ab}^{-1}$ integrated
luminosity are expected to be of similar size. For $50\, {\rm ab}^{-1}$
integrated luminosity the dominant  uncertainty will come from the $B \to \rho$
form factor. The sensitivity to New Physics is derived from the slope as a
function of $\epsilon_R$. For $A_{\rm FB}$ there is only a modest dependence,
that reduces significantly for positive admixture, reducing the sensitivity
considerably. 

\begin{figure*}[t]
\centerline{\includegraphics[width=.8\columnwidth]{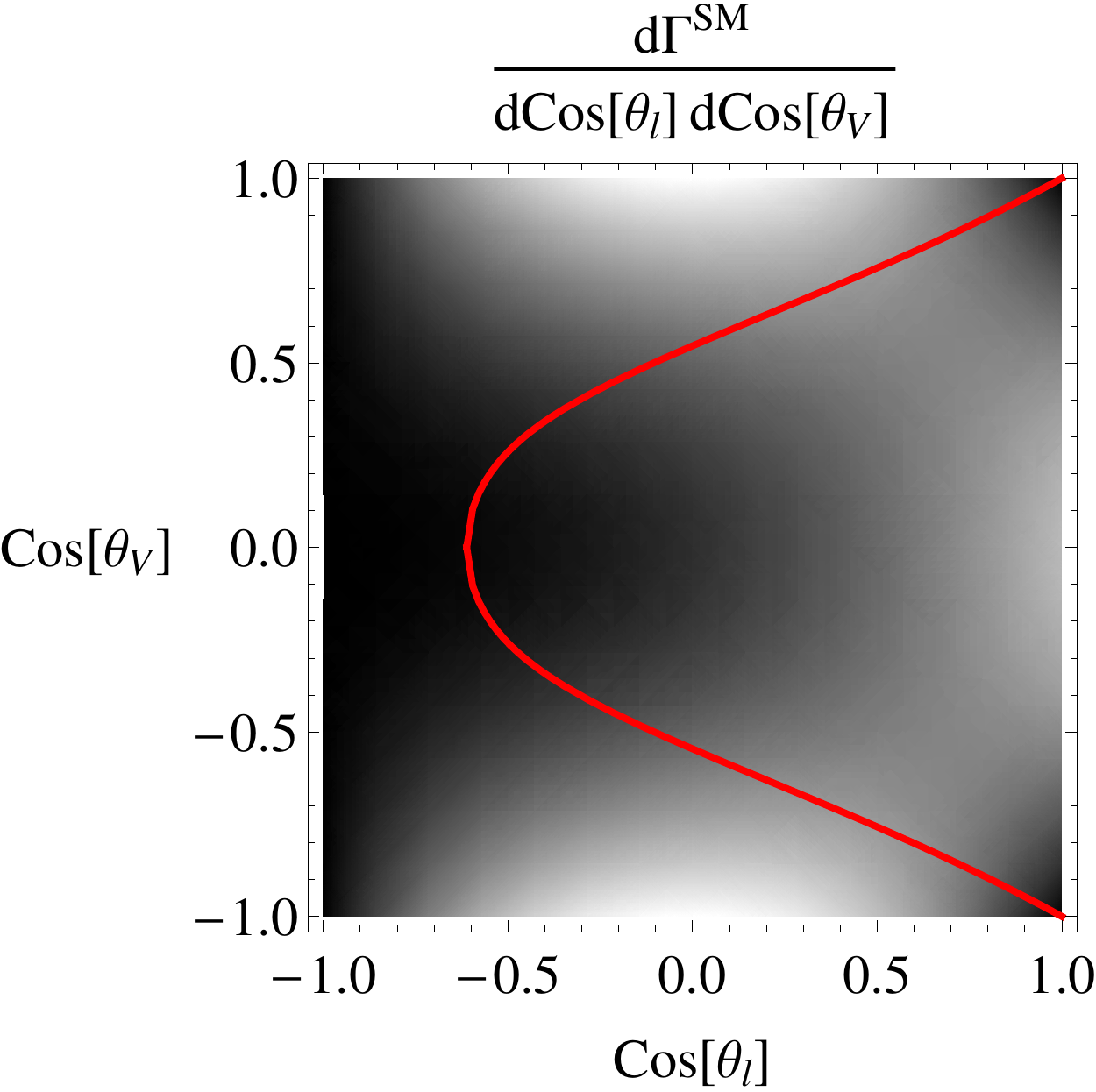} \hfil
\includegraphics[width=.8\columnwidth]{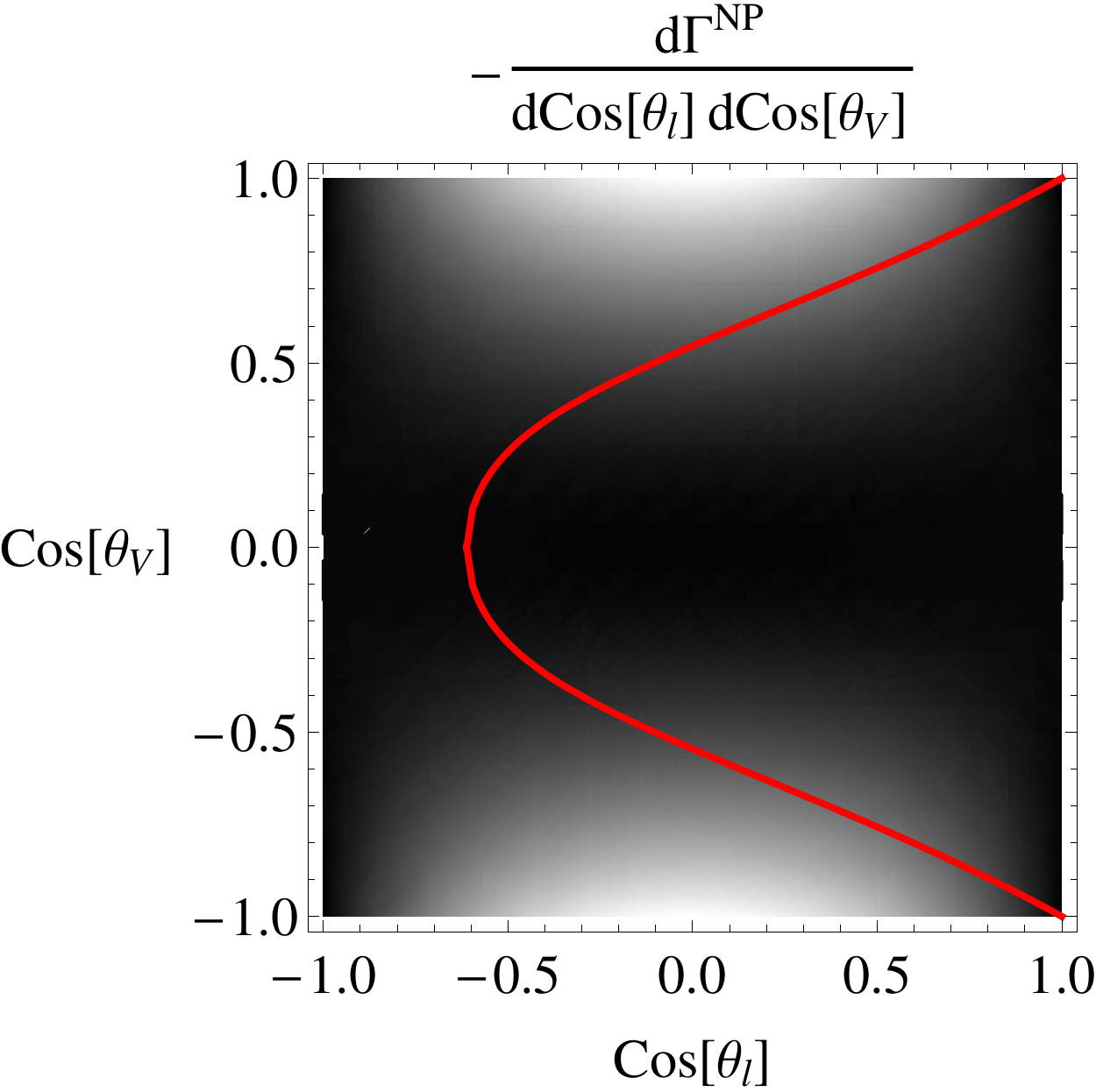}}
\caption{The optimized contour (red curve) separating the SM (left) and NP
linear in $\epsilon_R$ (right) contributions.}
\label{fig:epsr_contour}
\end{figure*}

The generalized two-dimensional asymmetry, $S$, requires the measurements of the decay angles
$\theta_{V,\ell}$. An optimal contour in terms of sensitivity to right-handed admixtures
 in these angles is devised as follows: The differential decay rate can be rewritten as
\begin{align}
\frac{\text{d}\Gamma}{\text{d}q^2\, \text{d}\cos\theta_V\, \text{d}\cos\theta_\ell}
= &\Big[f_\text{SM}^{(0)} (q^2,\cos \theta_\ell)  + \epsilon_R f_{\text{NP}_1}^{(0)} (q^2,\cos \theta_\ell) \nonumber \\
    &+ \epsilon_R^2 f_{\text{NP}_2}^{(0)} (q^2,\cos \theta_\ell) \Big] \nonumber\\
  + &\Big[f_\text{SM}^{(1)} (q^2,\cos \theta_\ell)  + \epsilon_R f_{\text{NP}_1}^{(1)} (q^2,\cos \theta_\ell) \nonumber \\
    +& \epsilon_R^2 f_{\text{NP}_2}^{(1)} (q^2,\cos \theta_\ell) \Big] \cos^2 \theta_V \,,\label{diff_rate}
\end{align}
where the functions $f^{(n)}_i$ are second order polynomials in $\cos
\theta_\ell$ and depend on $q^2$ explicitly, as well as indirectly through the
form factors
\begin{align}
    f^{(0)} &= J_{1s} + J_{2s} + J_{6s} \cos \theta_\ell - 2 J_{2s}
    \cos^2\theta_\ell \,,\nn\\
    f^{(1)} &= J_{1c} + J_{2c} -J_{1s}  -J_{2s} + (J_{6c}- J_{6s}) \cos \theta_\ell  \nonumber \\
    &\phantom{ = } + 2( J_{2s}-J_{2c}) \cos^2\theta_\ell\,.
\end{align}
Treating $\epsilon_R$ as a small parameter,
Eq.~\eqref{NewVariable} becomes
\begin{align}
  A &= A_\text{SM} + \epsilon_R\, A_{\text{NP}_1}
    + \epsilon_R^2\, A_{\text{NP}_2}\,, \nonumber \\
  B &= B_\text{SM} + \epsilon_R\, B_{\text{NP}_1}
    + \epsilon_R^2\, B_{\text{NP}_2}\,, \\
  S &= \frac{A_\text{SM}-B_\text{SM}}{A_\text{SM}+B_\text{SM}} 
    + 2 \epsilon_R\, \frac{A_{\text{NP}_1} B_\text{SM} -A_\text{SM} B_{\text{NP}_1}  }{(A_\text{SM}+B_\text{SM})^2}
    +\ldots\,. \nonumber
\end{align}
In the following we require $S_\text{SM} \approx 0$ what approximately divides the phase-space equally in the two regions of the asymmetry. The sensitivity to $\epsilon_R$ is optimized by demanding a maximal slope,
\begin{equation}
  \frac{\text{d} S}{\text{d} \epsilon_R} 
    = 2\, \frac{A_{\text{NP}_1} B_\text{SM} -A_\text{SM} B_{\text{NP}_1}}
    {(A_\text{SM}+B_\text{SM})^2} + {\cal O} (\epsilon_R)\,.
\end{equation}
This implies that
\begin{align}
    A_\text{SM} &\approx B_\text{SM}\,, \nn\\
    A_{\text{NP}_1} &\gg B_{\text{NP}_1} \quad \text{or} \quad A_{\text{NP}_1} 
    \ll B_{\text{NP}_1}\,.
\end{align}
In addition, the SM left-handed and the NP right-handed couplings are  odd and
even in $\cos \theta_\ell$, and both are symmetric in $\cos \theta_V$. The dividing curves are derived as follows.  The SM and NP
differential distribution are separated along a curve of constant ratio, causing a deviation in ration in the presence of non negligible right-handed admixture. 
The symmetry in $\cos \theta_V$ forces one region to be within $\pm \cos\theta_V
(\cos\theta_\ell)$ with $\cos\theta_V (\cos\theta_\ell^\text{min}) = 0$.
The first relation $\text{d}\Gamma_\text{SM} = \kappa\,
\text{d}\Gamma_\text{NP}^{(1)} $ using \eqref{diff_rate} is written as
\begin{align}
&\int_{\Delta q^2} \text{d}q^2 \big[ f_\text{SM}^{(0)} + f_\text{SM}^{(1)} \cos^2 \theta_V \big ] 
\nonumber\\ 
&= \int_{\Delta q^2} \text{d}q^2  \kappa 
  \big[f_{\text{NP}_1}^{(0)} + f_{\text{NP}_1}^{(1)} \cos^2 \theta_V \big]\,,
\end{align}
 where arguments for $(q^2,\cos \theta_\ell)$ were suppressed for the $f$ functions
for brevity. This implies
\beq
\cos^2 \theta_V (\cos\theta_\ell) = \frac{\int_{\Delta q^2} \big[\kappa f_{\text{NP}_1}^{(0)}  - f_\text{SM}^{(0)} \big] \text{d}q^2 }{\int_{\Delta q^2} \big[-\kappa f_{\text{NP}_1}^{(1)} + f_\text{SM}^{(1)} \big] \text{d}q^2 }\,.
\eeq
From this immediately follows
\beq
\label{shape}
\cos \theta_V^\text{min,max}(\cos\theta_\ell)
= \pm \sqrt{\frac{\int_{\Delta q^2} \big[\kappa f_{\text{NP}_1}^{(0)} - f_\text{SM}^{(0)} \big] \text{d}q^2 }{\int_{\Delta q^2} \big[-\kappa f_{\text{NP}_1}^{(1)}  + f_\text{SM}^{(1)} \big] \text{d}q^2 }}\,.
\eeq
The minimal value for $\cos\theta_\ell^\text{min}$ can be numerically obtained via
\begin{align}
    \int_{\Delta q^2} \text{d}q^2 \big[\kappa f_{\text{NP}_1}^{(0)} (q^2,\cos \theta_\ell^\text{min})\big]
        &=\int_{\Delta q^2} \text{d}q^2 \big[ f_\text{SM}^{(0)} (q^2,\cos \theta_\ell^\text{min})\big]  \,.
\end{align}
Note that the $f_i$ functions may be negative and thus the minimum value cannot
be imposed by having the integrand zero. Thus $\cos \theta_\ell^\text{min}$ will
depend on the interval $\Delta q^2$ and one numerically has $\cos
\theta_\ell^\text{min} = -0.611$. Independent of the actual form factor shape,
for $\cos \theta_V=1$ one has
\beq
\kappa f_{\text{NP}_1}^{(0)} - f_\text{SM}^{(0)} 
  = -\kappa f_{\text{NP}_1}^{(1)}  + f_\text{SM}^{(1)} \,,
\eeq
so that $\cos \theta_\ell^\text{max} =1$ and $\kappa$ is determined by requiring $S_\text{SM} \approx 0$.
The resulting curve most sensitive for $\epsilon_R$ that separates regions $A$ and $B$ can be numerically
approximated by
\begin{equation}
\cos \theta _{V} = \pm \sqrt{\frac{0.8472 \cos^2 \theta _\ell 
  +1.9038\cos\theta _\ell +0.8472}{-1.1484 \cos^2 \theta _\ell 
  +1.9038 \cos \theta _\ell +2.8429}} \,.
\end{equation}
This choice depends on nonperturbative input quantities. However, it turns out
that in the heavy quark limit the minimum value $\cos \theta_\ell^\text{min}$ is
given form-factor independently and agrees well with the one derived from the
full form factors, while the shape $\cos \theta_V$ ($\cos\theta_\ell$) is
distorted mildly using the heavy quark limit form factors. This curve is
displayed together with the SM and NP density distributions in
Fig.~\ref{fig:epsr_contour}.

The Neyman belt of $S$ and sensitivities are shown in
Fig.~\ref{fig:SE_lin_AFB_S}: integrating over a range of $q^2$ introduces
additional uncertainties, that do not cancel entirely in the ratio, resulting in
larger theoretical uncertainties than for $A_{\rm FB}$. The overall sensitivity
on NP for $1\,{\rm ab}^{-1}$ of integrated luminosity, however, is better due to
the increased dependence on $\epsilon_R$, and for $50\,{\rm ab}^{-1}$ of data
the sensitivity is comparable. 

\subsection{Simple generalized ratios}

\begin{figure*}[t]
\centerline{
\includegraphics[width=.65\columnwidth]{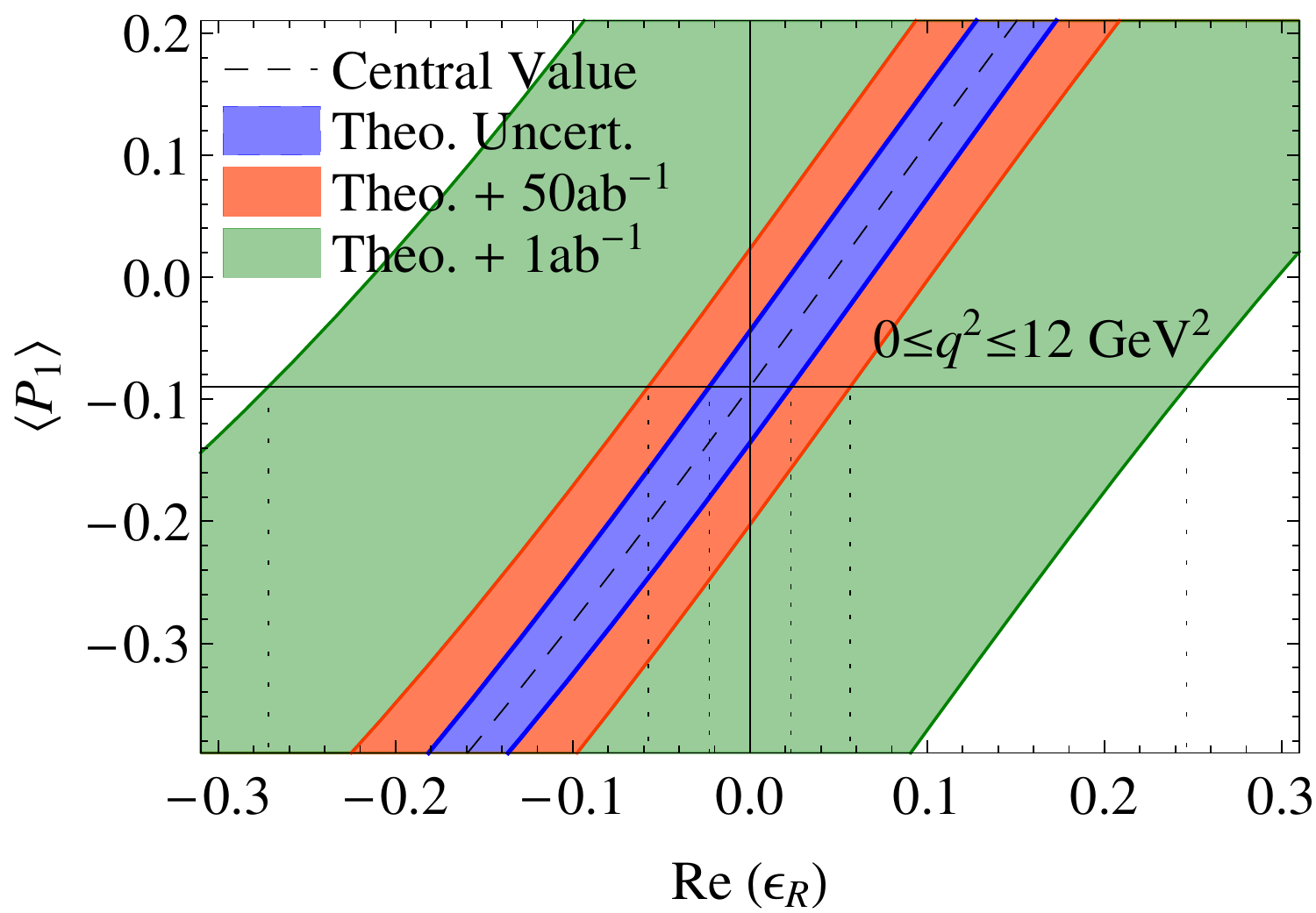}\hfill
\includegraphics[width=.65\columnwidth]{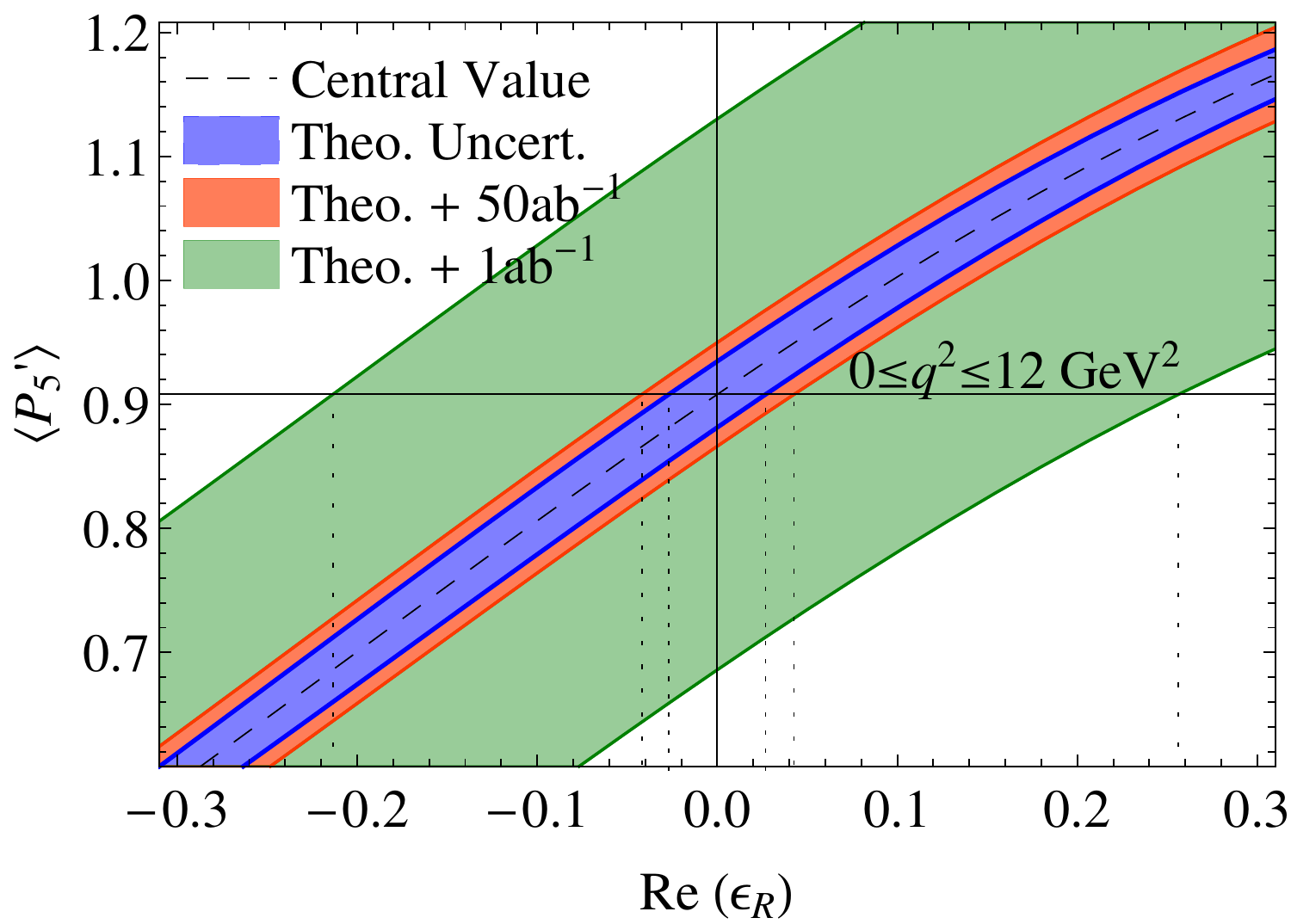}\hfill
\includegraphics[width=.65\columnwidth]{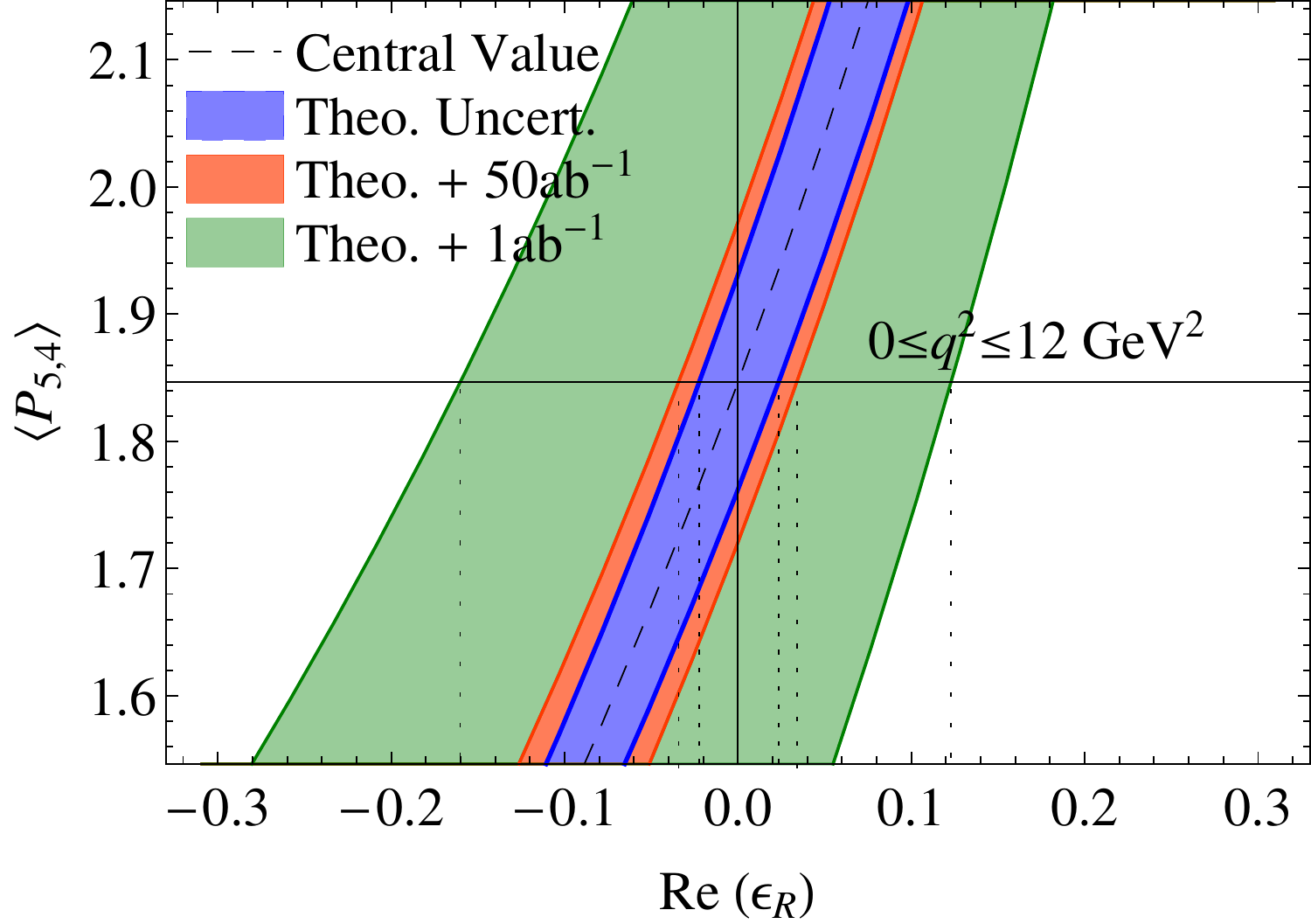}}  
\centerline{
\includegraphics[width=.65\columnwidth]{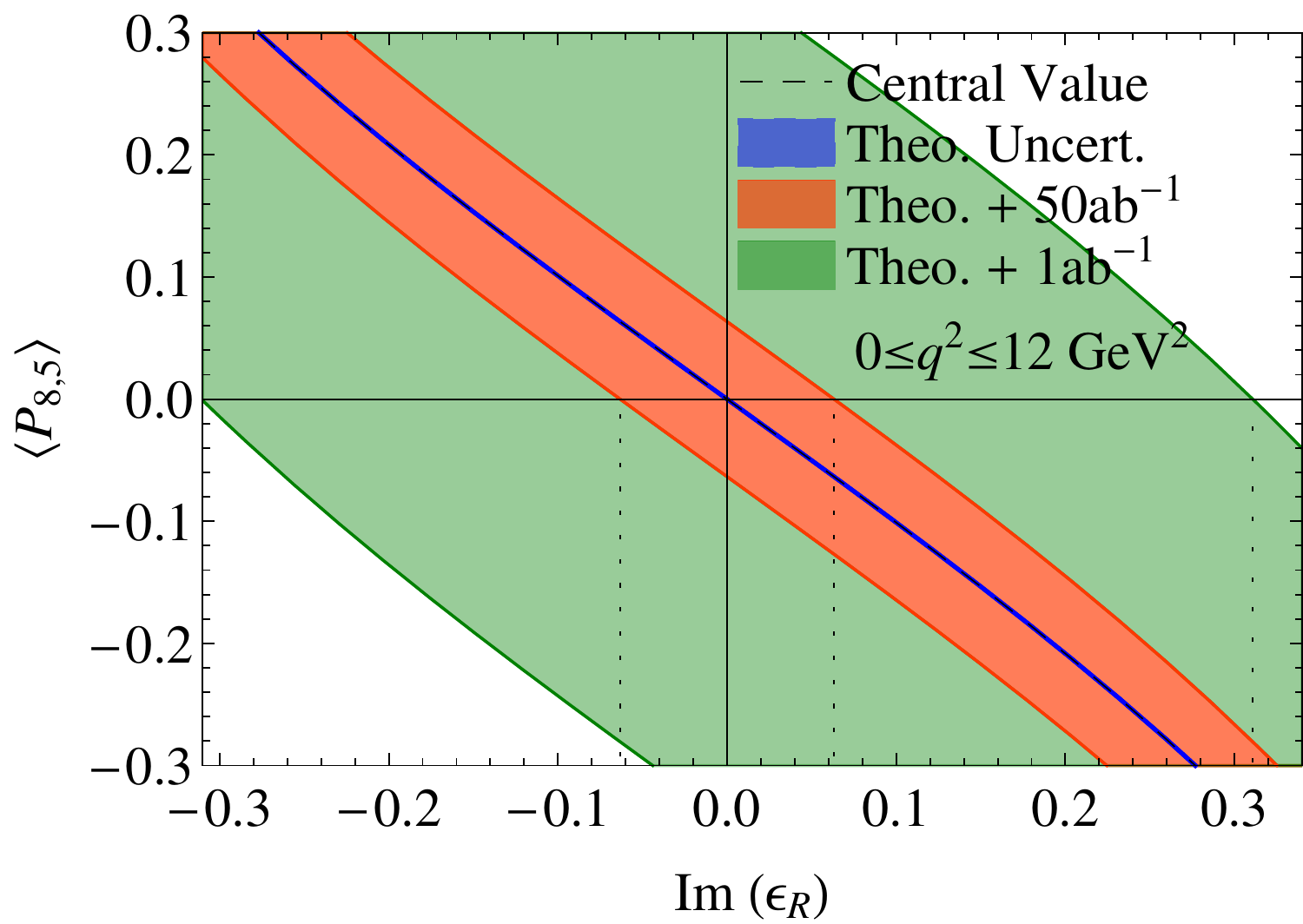}\hfill
\includegraphics[width=.65\columnwidth]{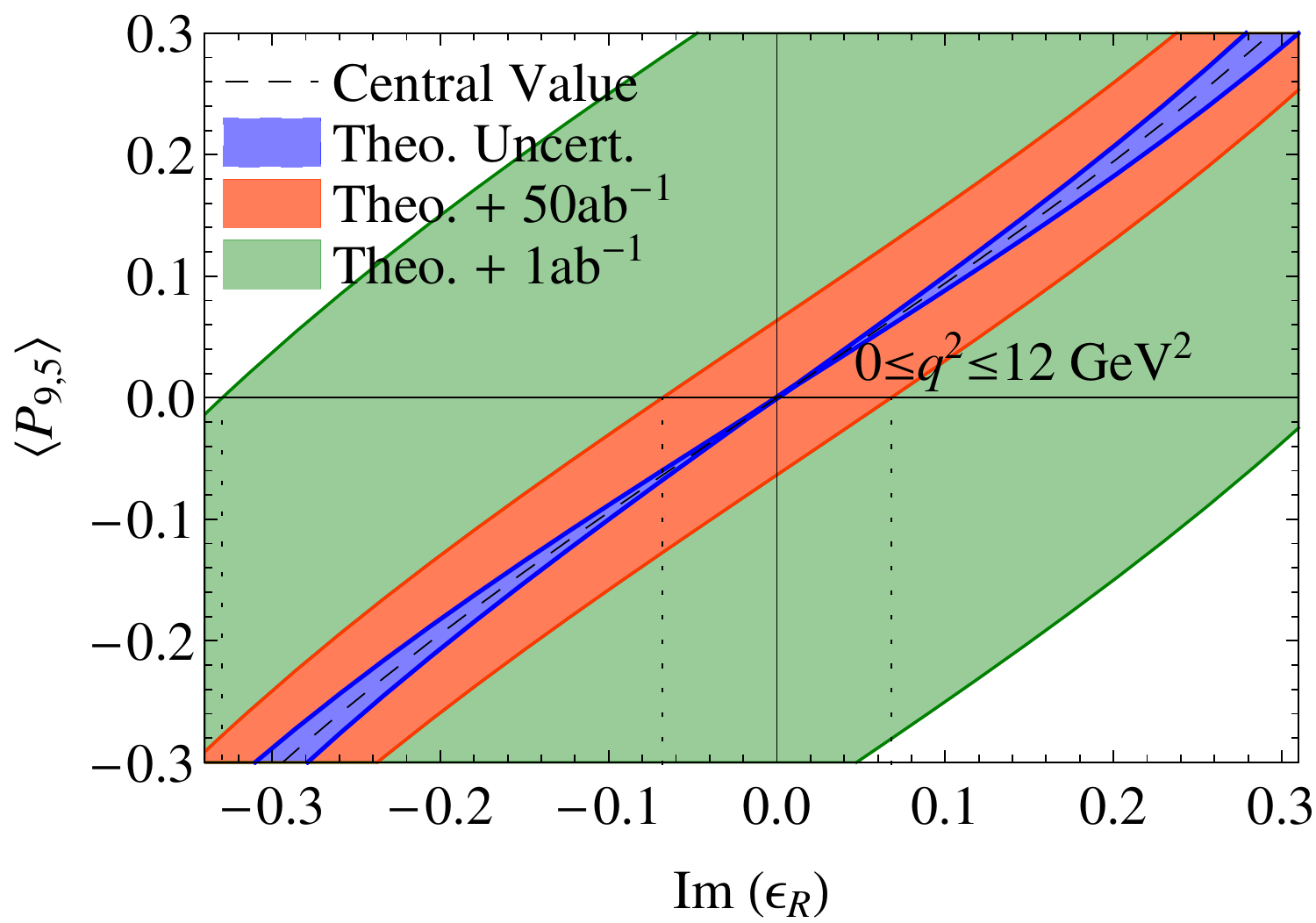}\hfill
\includegraphics[width=.65\columnwidth]{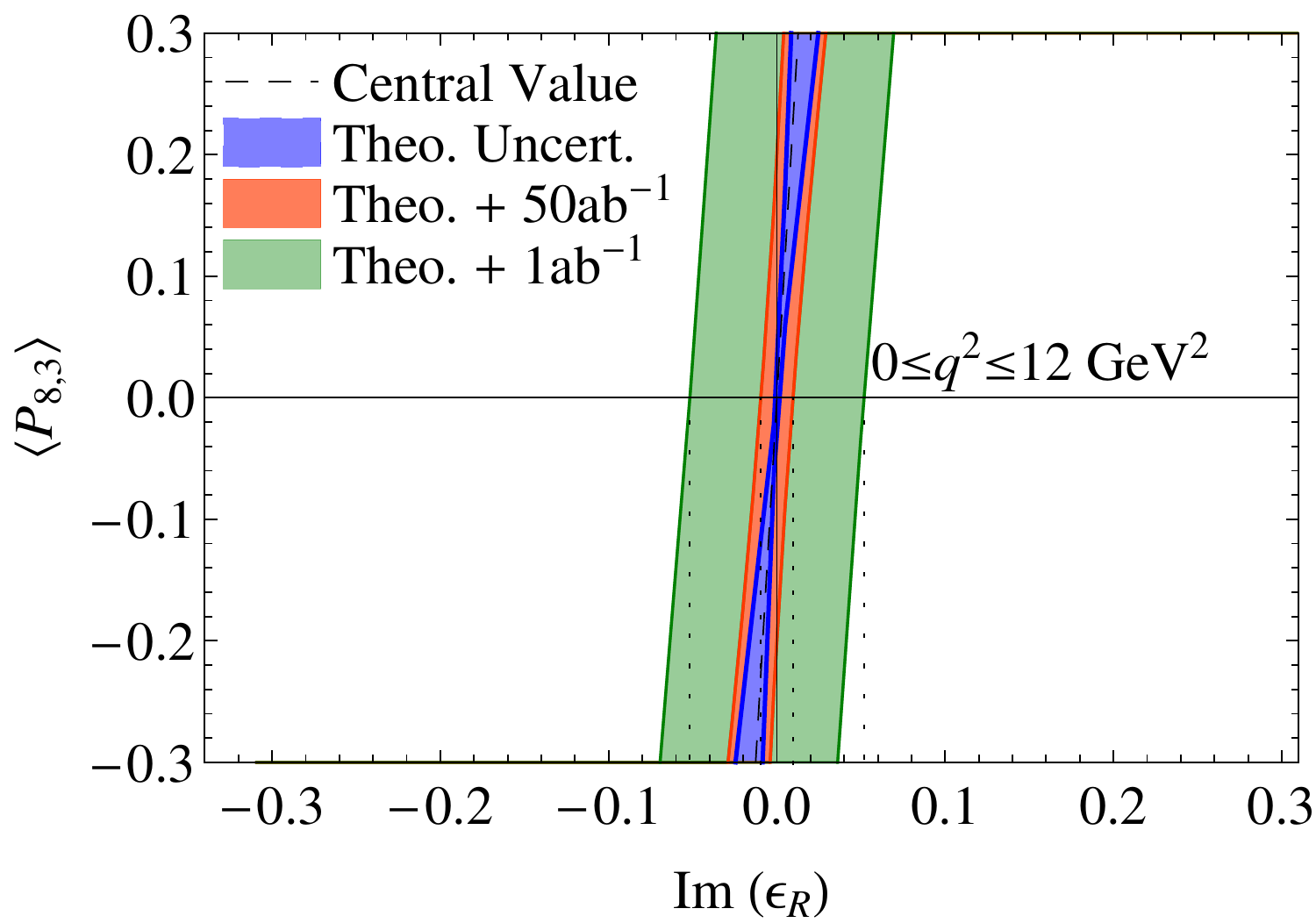}}
\caption{The most sensitive angular observables to ${\rm Re}\,\epsilon_R$ (top
row) and to ${\rm Im}\,\epsilon_R$ (lower row), assuming $\epsilon_R$ to be
purely real or imaginary, respectively. The blue bands show the theoretical
uncertainties, while the orange [dark-green] bands show theory and experimental
uncertainties combined in quadrature, for 50\,ab$^{-1}$ [1\,ab$^{-1}$] of
$B$-factory data. The observables, $\langle P_1\rangle$ (top left), $\langle
P_5'\rangle$ (top center),  $\langle P_{5,4}\rangle$ (top right), $\langle
P_{8,5}\rangle$ (lower left), $\langle P_{9,5}\rangle$ (lower center), $\langle
P_{8,3}\rangle$ (lower right), are defined in
Eqs.~\eqref{clean_sens1}--\eqref{clean_sens4}.}
\label{opt_sensitive}
\end{figure*}

In the context of ``clean observables'' a set of simple generalized observables,
$P_i$, in Eq.~(\ref{clean_sens1}-\ref{clean_sens4}), from which one expects the best
theoretical sensitivity are derived. The most sensitive observables in the context of real right-handed currents,
are $\langle P_1 \rangle$, $\langle P_{5}'\rangle$ and $\langle P_{5,4}\rangle$.
The most sensitive observables for imaginary right-handed contributions are 
$\langle P_{8,5} \rangle$, $\langle P_{9,5}\rangle$ and $\langle  P_{8,3}\rangle$. The corresponding predictions and sensitivities are shown in Fig.~\ref{opt_sensitive}. 
The statistical correlations between the numerator and denominator in the observables was estimated using Monte Carlo methods, neglecting any influence from background.
The three-dimensional observables reduce the theoretical uncertainties with respect
to the one-dimensional or two-dimensional asymmetries. 
Their experimental uncertainties, however, are larger due
to the great number of free parameters that need to be determined from the same data. The most precise observable
for $1\,{\rm ab}^{-1}$ of integrated luminosity are $\langle P_{5,4}\rangle$ and $\langle P_{8,3}\rangle$,
for real and imaginary $\epsilon_R$, respectively. 

\subsection{Testing NP contributions vs.\ form factor uncertainties}

The predicted value of the observables depends on the assumed form factor shape
and integrated $q^2$ range. As this is a nonperturbative calculation with
possible unknown systematic uncertainties, in case experimentally a significant
deviation is observed, it is necessary to verify if NP is the source of a
possible deviation (see the recent discussion related to $B\to K^*$
transitions~\cite{Jager:2012uw}).

An obvious consistency check is to measure several of the presented
observables.  In addition one should perform a $q^2$ binned analysis of these,
for instance reconstruct them in a high $q^2$ and a low $q^2$ range.  If a
measured deviation is related to not properly considered theoretical or also
experimental uncertainties, it will produce an inconsistent pattern, since one
expects all  regions in $q^2$ to show a consistent deviation form the SM due to
a right-handed admixture.

In addition two of the ``clean observables", $\langle P_4 \rangle$ and $\langle
P_5 \rangle$, are nearly independent of a right-handed current,
\begin{align}
   \langle P_4\rangle_\text{bin} &= \frac{\sqrt{2}\int_{\Delta q^2} \text{d}q^2 J_4}{\sqrt{- \int_{\Delta q^2} \text{d}q^2 J_{2c}\,\,\int_{\Delta q^2} \text{d}q^2 (2 J_{2s} - J_3)   }}\,,\nonumber \\
    &\approx 0.94 \pm 0.01_\text{Theory}\,, \label{clean_insens1}\\
   \langle P_5 \rangle_\text{bin} &= \frac{\int_{\Delta q^2} \text{d}q^2 J_5}{\sqrt{- \int_{\Delta q^2} \text{d}q^2 2 J_{2c}\,\,\int_{\Delta q^2} \text{d}q^2 (2 J_{2s} + J_3   )}}\nonumber \\
    &\approx 0.95 \pm 0.01_\text{Theory}\,. \label{clean_insens2}
\end{align}
Thus a global hypothesis test incorporating all observables, taking into account
the proper experimental and theoretical correlations, would be desirable and
be the most powerful test of the data for the presence of right-handed
currents.

Note that $\langle P_4 \rangle$ is also insensitive to ${\rm Im}\,\epsilon_R$,
while there is a quadratic effect in $\langle P_5 \rangle$. This happens to be a
coincidence in cancellation of the NP parameters in the point-by-point ratio,
which is broken by the finite binning. However, numerically this breaking amounts to
a very small, unobservable effect. In case these $P_i$ can be
measured using for instance the asymmetries in Table.~\ref{tab:asym}, one can
use this prediction to test for other effects.

\section{\boldmath Global fit for $|V_{ub}^L|$ and $\epsilon_R$}
\label{sec:global_fit}

The estimated sensitivities on $\epsilon_R$ in the previous section can be used
to add an orthogonal  constraint to the global fit  performed in
Section~\ref{sec:intro}.  The gain in overall sensitivity on $|V_{ub}^L|$ and
$\epsilon_R$ is estimated by extrapolating  the experimental uncertainties to $1
\, {\rm ab}^{-1}$ and $50 \, {\rm ab}^{-1}$. For the branching fraction input
other than $B \to \rho \ell \bar \nu$ and $B \to \pi \ell \bar \nu$ the
projections from Ref.~\cite{Aushev:2010bq}  are used. For $B\to \pi \, \ell
\bar\nu$ a more optimistic uncertainty of $3\%$ is used due to estimated
progress in the lattice QCD form factor determinations~\cite{USlqcd}. For the $B
\to \rho \ell \bar\nu$ branching fraction the uncertainties discussed in
Section~\ref{sec:observables} are used and are listed in
Table~\ref{tab:demo_Bfit_simple}. The irreducible uncertainty of the $B \to \rho
\, \ell \bar\nu$ form factors is quoted to be 7\% in Ref.~\cite{Ball:2004rg}. In
the following no such scenarios are explored, due to the complication related to
how a reduction of uncertainty would affect the overall correlations between the
different form factors. 

\begin{figure*}[t]
\centerline{
\includegraphics[width=.9\columnwidth]{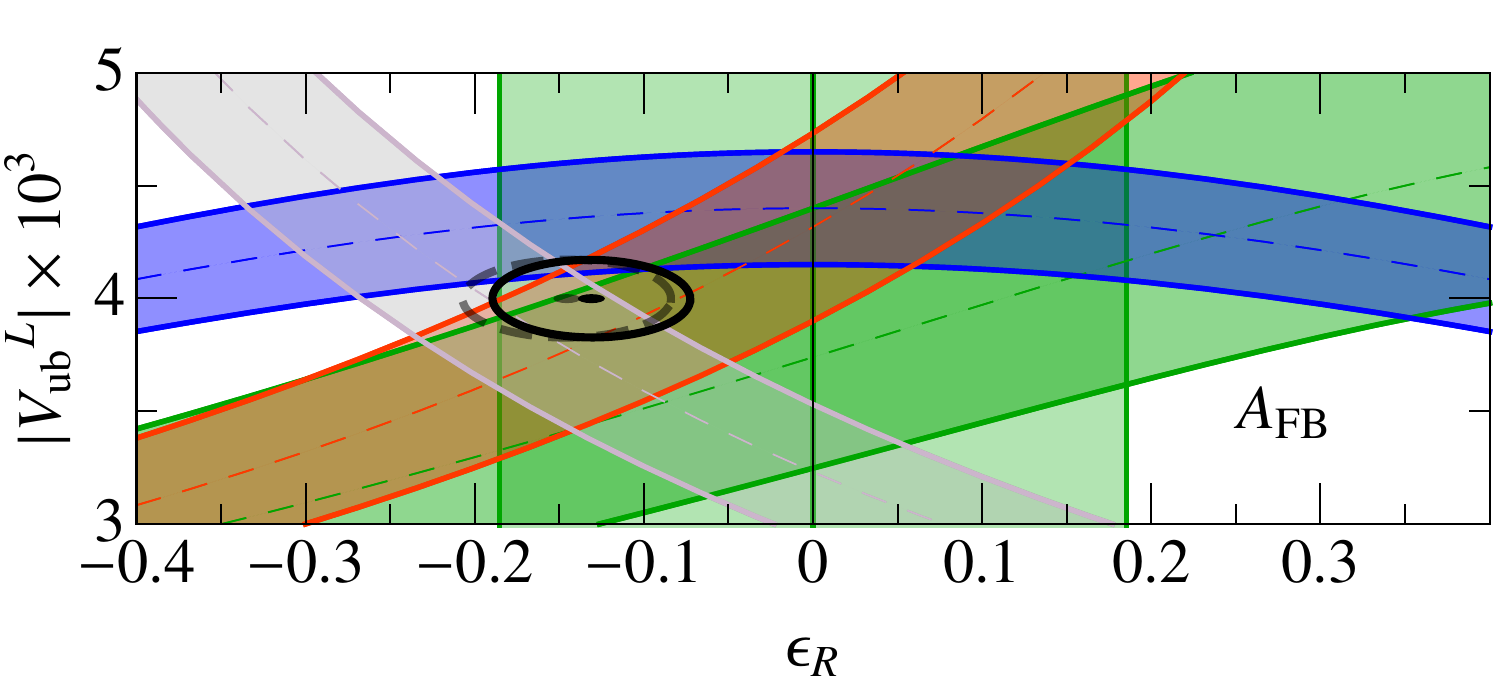} \hfil
\includegraphics[width=.9\columnwidth]{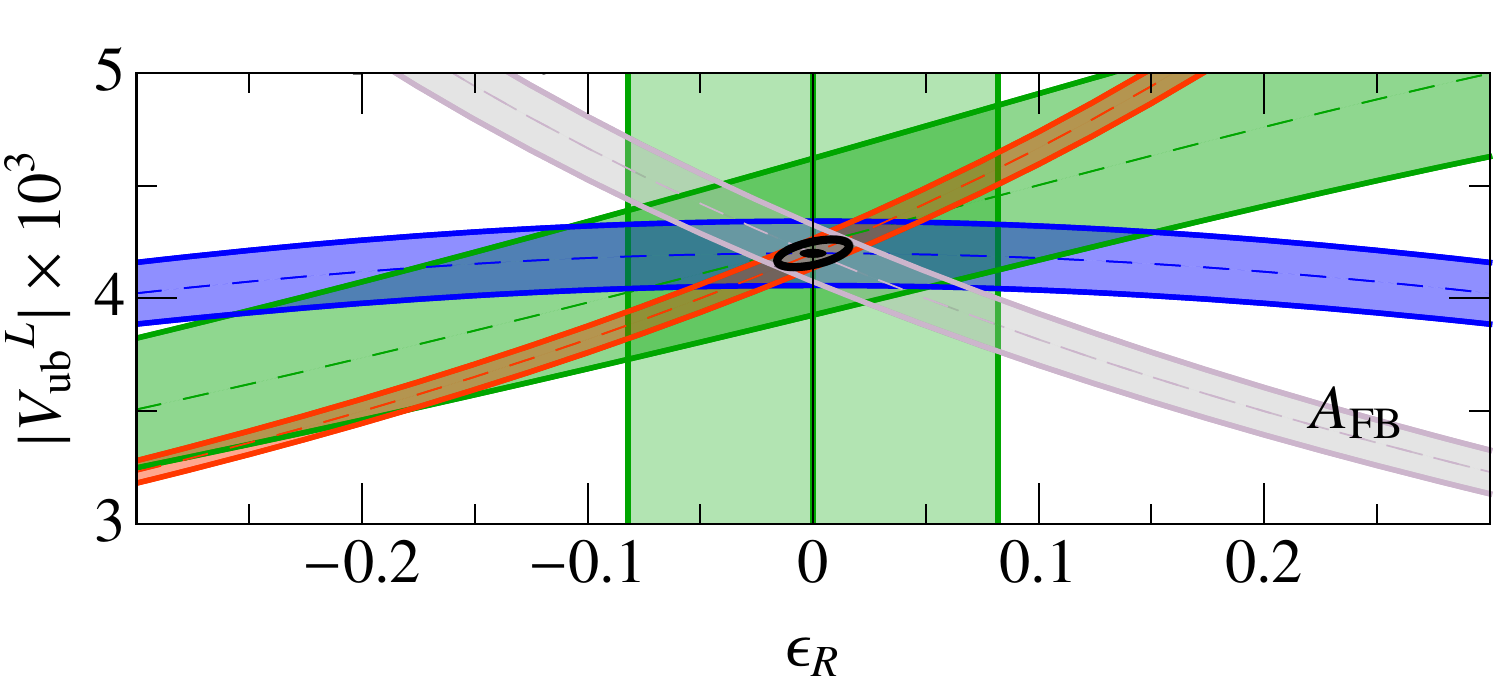} 
}
\centerline{
\includegraphics[width=.9\columnwidth]{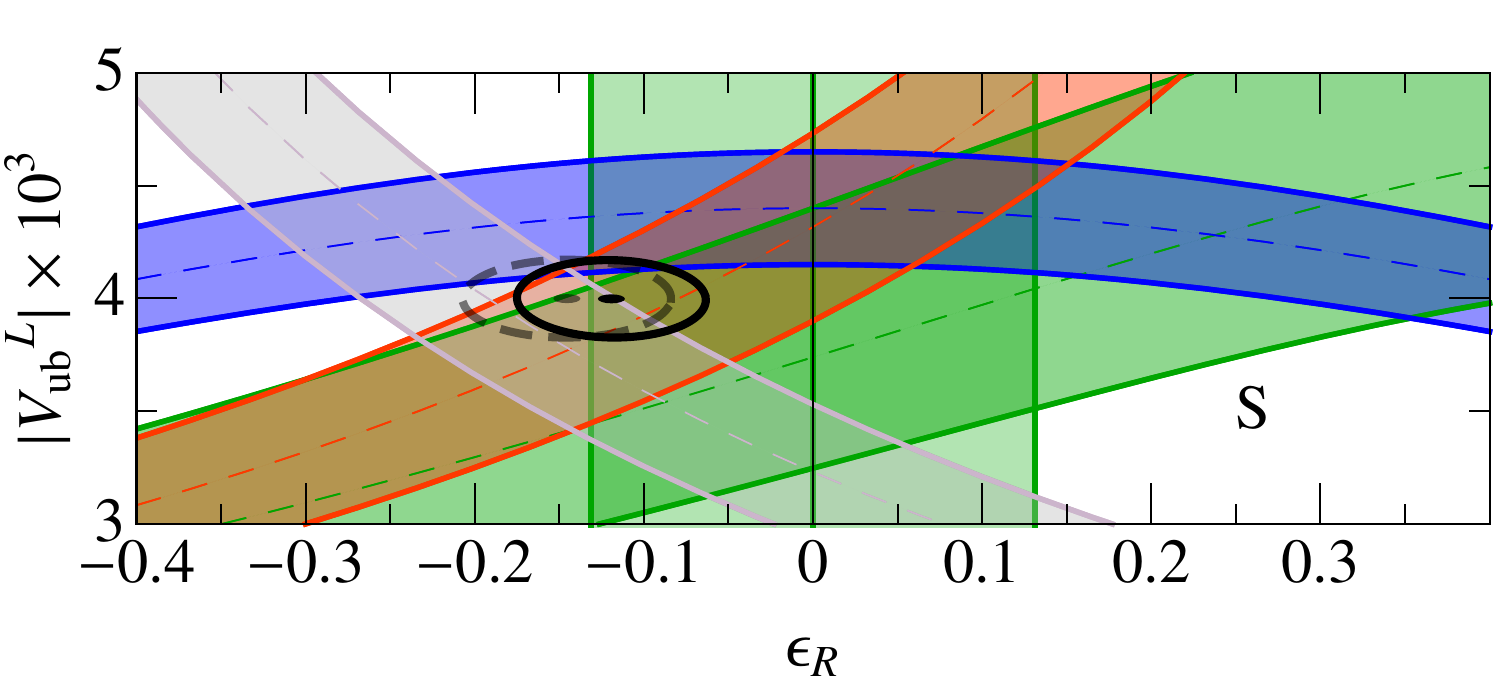} \hfil
\includegraphics[width=.9\columnwidth]{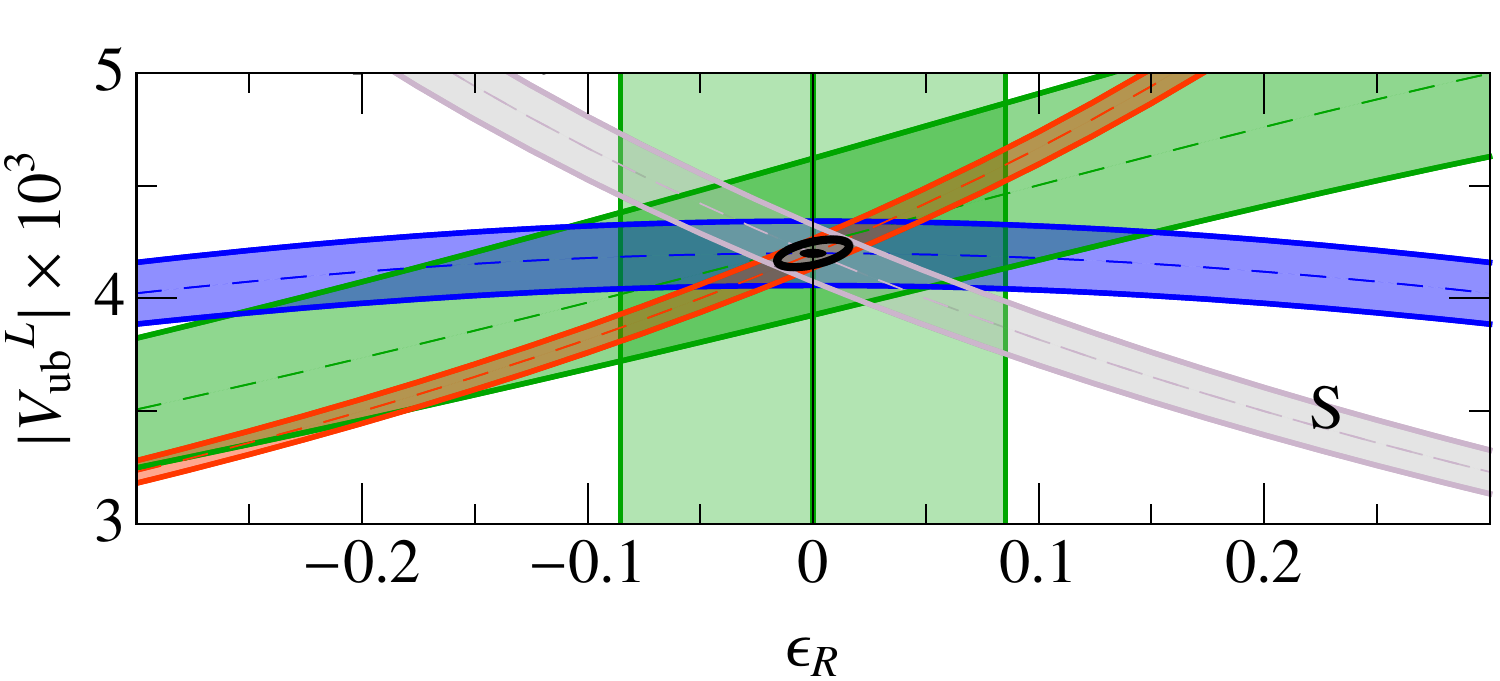} 
}
\centerline{
\includegraphics[width=.9\columnwidth]{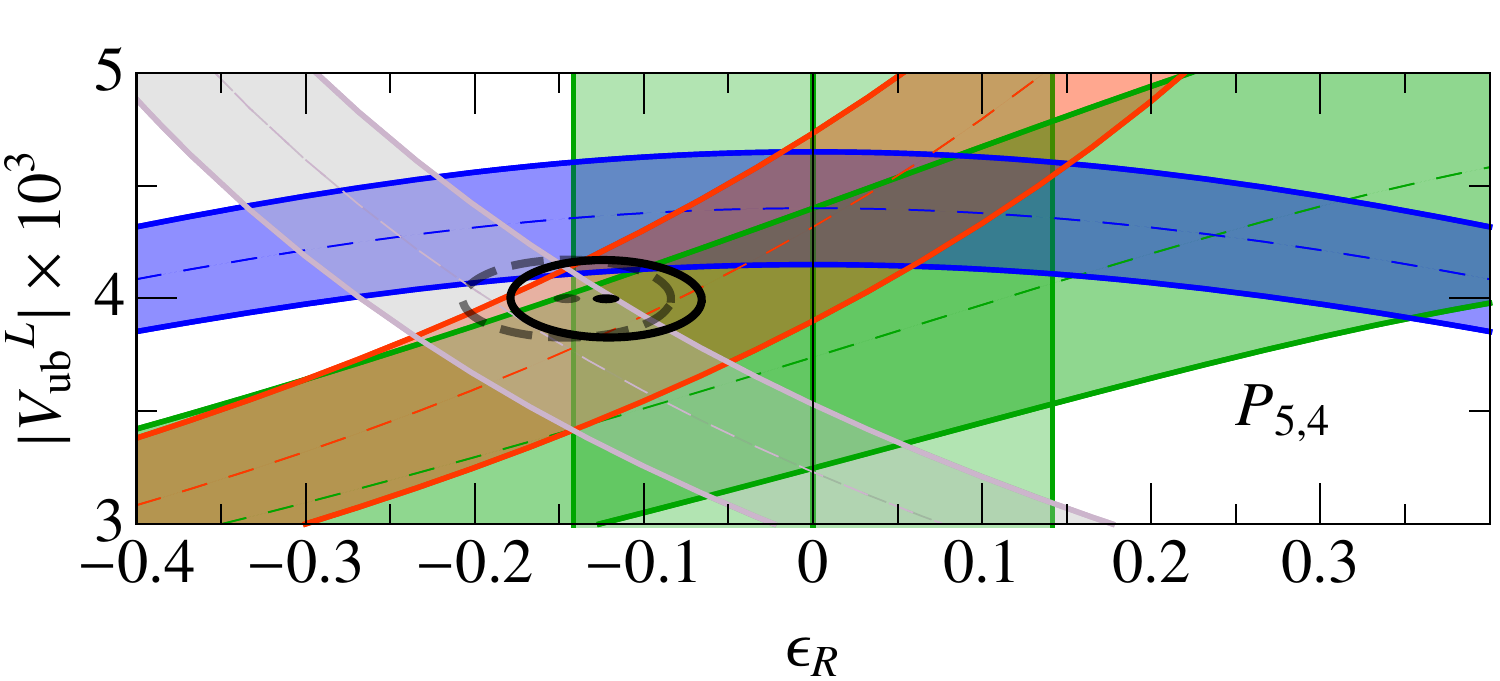}\hfil
\includegraphics[width=.9\columnwidth]{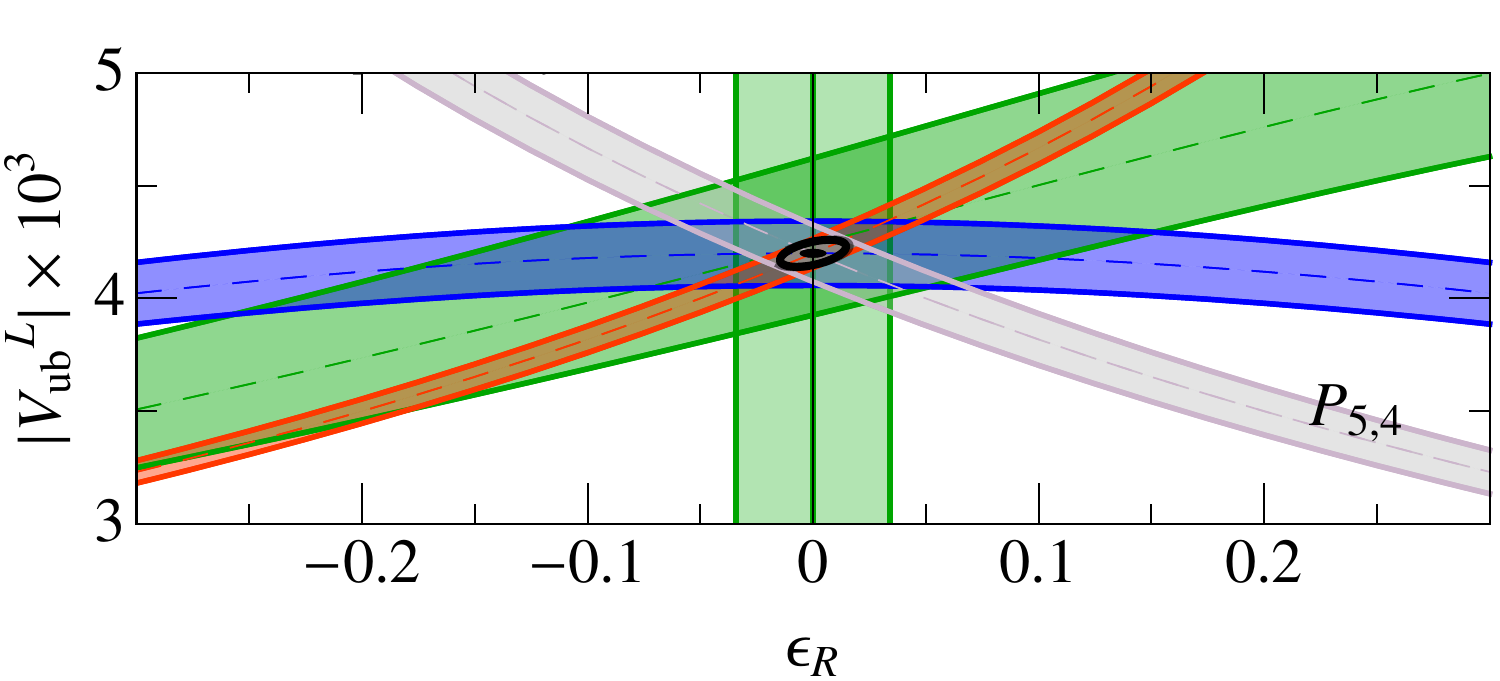}
}
\caption{The $\chi^2$ fits for $|V_{ub}^L|$ and $\epsilon_R$ assuming $1\, {\rm
ab}^{-1}$ (left) and $50\, {\rm ab}^{-1}$ (right) of $B$-factory data.  The
green bands show the $B \to \rho \ell \bar \nu$ information, c.f.,
Fig.~\ref{fig:epsr_chisqfit}. The observable used for the expected orthogonal
bound on $\epsilon_R$, assuming the SM, is shown in each Figure. The used
uncertainties for $50 \, {\rm ab}^{-1}$ are quoted in
Tables~\ref{tab:demo_Bfit_simple}. Table~\ref{tab:fit_Vubl_details} lists the
improvement in uncertainty by including the orthogonal constraint from the
discussed observable on $\epsilon_R$ with respect to the uncertainty of fitting
the experimental information available by $B \to X_u \ell \bar \nu$, $B \to \tau
\bar \nu$, $B \to \pi \ell \bar \nu$, and $B \to \rho \ell \bar \nu$ only.}
\label{fig:demo_fit}
\end{figure*}

Fig.~\ref{fig:demo_fit} shows the results for the simultaneous fit for
$|V_{ub}^L|$ and $\epsilon_R$ for integrated luminosities of $1 \,{\rm ab}^{-1}$
and $50 \, {\rm ab}^{-1}$.  The fits incorporate the expected constraints from
either $A_{\text{\rm FB}}$, $S$, or $P_{5,4}'$ in the absence of right-handed
currents. For the $1\,{\rm ab}^{-1}$ scenario, the current experimental central
values are used for $|V_{ub}|$, whereas for $50\, {\rm ab}^{-1}$ the SM is
assumed, with identical $|V_{ub}|$ from all channels. For $1\,{\rm ab}^{-1}$
$B$-factory data, $S$ results in the largest gain in sensitivity for
right-handed currents among the studied observables. 
Table~\ref{tab:fit_Vubl_details} lists the reduction of the uncertainty of
$|V_{ub}^L|$ and $\epsilon_R$ with respect to a fit without any additional
orthogonal bound. Although the theoretical uncertainties on $S$ are more sizable
than on $P_{5,4}'$, the experimental simplicity of the two-dimensional
asymmetry results in the best overall expected sensitivity. The reduction in
experimental uncertainties for $50\,{\rm ab}^{-1}$ statistics changes this
picture: here the theoretical uncertainties on the $B \to \rho$ form factors
dominate the overall uncertainty of all observables and $P_{5,4}'$ results in
the best expected sensitivity. 

\begin{table}[b]
\begin{tabular}{ccccc}
\hline\hline
Decay  & Expected error on $|V_{ub}|$  \\
\hline
$B \to \pi \, \ell \bar\nu$ & 3\% \\
$B \to X_u \ell \bar\nu$  & 3\% \\
$B \to \tau \, \bar \nu_\tau$  & 1.5\% \\
\hline\hline
Decay  & Expected error on $\mathcal{B}$  \\ \hline
$B \to \rho \, \ell \bar\nu$ & 3\% \\
\hline\hline
\end{tabular}
\caption{The assumed uncertainties on $|V_{ub}|$ and $\mathcal{B}(B \to \rho
\ell \bar \nu$  for $50 \, {\rm ab}^{-1}$ of integrated luminosity are listed.}
\label{tab:demo_Bfit_simple}
\end{table}

\begin{table}[bth]
\begin{tabular}{ccc}
\hline\hline
Fit  &  $\delta \left( \left| V_{ub}^L \right| \right)$ [\%] & $\delta \left( \epsilon_R \right)$ [\%] \\
\hline
4 modes + $A_{\text{FB}}$ (1 ab${}^{-1}$)  & $-0.3$& $-5$   \\
4 modes + $S$ (1 ab${}^{-1}$)  & $-0.5$ & $-9$   \\
4 modes + $P_{5,4}$ (1 ab${}^{-1}$)  & $-0.5$ & $-8$   \\
\hline
4 modes + $A_{\text{FB}}$ (50 ab${}^{-1}$)  & $-0.4$ & $-2$   \\
4 modes + $S$  (50 ab${}^{-1}$)  & $-0.5$ & $-2$ \\
4 modes + $P_{5,4}$ (50 ab${}^{-1}$)  & $-3$ & $-10$  \\
\hline\hline
\end{tabular}
\caption{The expected relative reduction in the uncertainty of $\left| V_{ub}^L
\right| $ and $\epsilon_R$ for the $\chi^2$ fits in Figs.~\ref{fig:demo_fit}.
The improvements are quoted with respect to the expected uncertainties on the
4-mode analysis for 1 ab${}^{-1}$ and 50 ab${}^{-1}$, which are $\Delta \left(
\left| V_{ub}^L \right| \times 10^3, \Delta \epsilon_R \right) = \left( 0.18,
0.061 \right)$ and $  \left( 0.06, 0.016 \right)$, respectively.}
\label{tab:fit_Vubl_details}
\end{table}

\section{Summary and Conclusions}
\label{sec:conclusions}

In this paper, the full decay distribution in semileptonic $B\to \rho
[\rightarrow \pi \pi] \ell \bar{\nu}$ decay was analyzed to explore the
consequences of a possible right-handed semileptonic current from physics beyond
the Standard Model.  A number of observables was explored, some new and some
defined in the literature, and a detailed investigation of the impact of the
theoretical uncertainties on the sensitivity was performed.

The theoretical uncertainties and correlations are crucial ingredients of
predicting the uncertainties of the observables reliably. Such correlation
information is not readily available in existing $B \to \rho$  or other form
factor calculations. A model of these correlations is discussed for the $B \to
\rho$ form factor, incorporating correlations among different form factors and
different $q^2$ points of the form factors

A detailed analysis on the $B \to \rho$ form factor was performed. The
use of unitarity constraints to predict the form factor shape up to a small
expansion  was revisited and verified In the context of a vector meson final
state.  This technique is known in the literature as $z$-parametrization, and
has the advantage to be valid over the entire $q^2$ range of the form factor. In
order to combine all this information, a fit routine was developed, which is
capable of fitting several correlated or uncorrelated sources of form factor
values, i.e., LCSR and lattice QCD, taking into account the correlations among
the form factors and among different points of $q^2$. 
The fit results have been cross-checked with several parametrizations and
validated by fits of pseudo-input.

With the theoretical prediction for the fully differential spectrum including
correlations at hand, the sensitivity to a right-handed current was
investigated, which has been proposed as a possibility to ease a current
tensions in the determinations of $|V_{ub}|$. To set a bound on this beyond
Standard Model contribution, two approaches are possible: (i) a full
four-dimensional fit for the $J_i$ coefficients or counting experiments that
involve determining the partial branching fraction in several regions of phase
space and combining this information appropriately to project out either the
$J_i$ coefficients, or (ii) to construct asymmetries sensitive to NP
contributions. The latter offer an obvious alternative, since with the currently
available $B$-factory data, a full four-dimensional fit appears to be a very
challenging endeavor. 

The discussed observables exhibit very different theoretical and experimental
uncertainties: besides the usual forward-backward asymmetry, a two-dimensional
generalized asymmetry  is proposed by integrating out one of the decay angles
form the fully differential decay rate. These two are experimentally the
simplest observables. A set of generalized three-dimensional observables is
discussed. These are experimentally more challenging, and the eventual
observables involve ratios of statistically and systematically correlated
observables. 

A ranking in terms of sensitivity reveals that the balance of experimental and
theoretical  uncertainties is important: for the available $B$-factory
statistics of about $1\, {\rm ab}^{-1}$, the two dimensional asymmetry $S$ with
its simple experimental definition seems to be the most sensitive to the
presence of right-handed currents.  For the anticipated $50\, {\rm ab}^{-1}$
Belle~II statistics, the more complicated three-dimensional observables result
in the best expected sensitivity due to the  reduction of experimental
uncertainties. A direct determination of $\epsilon_R$ allows to introduce an
orthogonal constraint into the indirect determination involving $\left| V_{ub}
\right|$ measurements from various decays with  different $\epsilon_R$
dependencies. Including the most sensitive direct $\epsilon_R$ constraint for
$1\, {\rm ab}^{-1}$ or $50\, {\rm ab}^{-1}$, reduces the uncertainty of
$\epsilon_R$ by about 10\% in such a global analysis. This implies that even
with the current $B$-factory  datasets a useful statement about $\epsilon_R$
from $B \to \rho \ell \nu$ can be obtained.

In case a deviation from the SM is observed, a global hypothesis test
incorporating all observables is desirable: the presence of a right-handed
admixture should result in a consistent change. A non-consistency could imply
problems with the predictions of the $B \to \rho$ form factors.  To assert the
correctness of the form factor predictions, an analysis in bins of $q^2$ (e.g.,
a low and high $q^2$ region) should be performed to see if the deviation is 
consistent and independent of the $q^2$ region. Ultimately a fully differential
analysis  by a four dimensional fit would be desirable to analyze the full
anatomy of this decay mode. In the context of such an analysis, the
nonperturbative uncertainties would be greatly reduced. 

Future high statistics $B\to\rho\ell\bar\nu$ measurements will not only allow
the extraction of the $J_i$ coefficients in this semileptonic decay and more
sensitive searches for right-handed current, it can also be used to test the
form factor relations, which are important for the interpretation of the
identically defined observables in $B\to K^*\ell^+\ell^-$.  If precise and
reliable lattice QCD calculations of $B\to\rho$ and $B\to K^*$ form factors
become available, the quoted theoretical uncertainties can be greatly reduced.
Such additional inputs can be readily incorporated into our fit for the form
factors and the combined analysis with the available experimental data. 
Furthermore, precise form factor input at high $q^2$ would allow to access the
whole kinematic region, increasing the statistical power of the experimental
data and hence improving the sensitivity to new physics in
$B\to\rho\ell\bar\nu$. 

\acknowledgments

We thank Wolfgang Altmannshofer, Danny van Dyk, Jernej Kamenik, Roman Zwicky, and Bob Kowalewski
for useful discussions.
The work of ZL was supported in part by the Office of Science, Office of High
Energy Physics, of the U.S.\ Department of Energy under contract
DE-AC02-05CH11231.  ZL thanks the hospitality of the Aspen Center for Physics
(NSF Grant PHY-1066293), where part of this work was carried out.
ST was supported by a DFG Forschungsstipendium under contract no.~TU350/1-1, by
the ERC Advanced Grant EFT4LHC of the European Research Council and the Cluster
of Excellence Precision Physics, Fundamental Interactions and Structure of
Matter (PRISMA-EXC 1098).

\appendix

\section{Partially Integrated Angular Rates}
\label{sec:app_partial}

The differential rates integrated over one angle are then given by
\begin{widetext}
\begin{align}
\frac{\text{d}\Gamma}{\text{d}\cos\theta_V\, \text{d}\cos\theta_\ell} 
  &= \frac{G_F^2 |V_{ub}^L|^2 m_B^3 }{ \pi^3}  \bigg\{ \bar J_{1s}  \sin^2 \theta_V + \bar J_{1c} \cos^2 \theta_V
  + ( \bar J_{2s} \sin^2 \theta_V + \bar J_{2c} \cos^2 \theta_V) \cos 2\theta_\ell \nn\\
& + ( \bar J_{6s} \sin^2 \theta_V  + \bar J_{6c} \cos^2 \theta_V ) \cos \theta_\ell \bigg\}\,,\\
\frac{\text{d}\Gamma}{\text{d}\cos\theta_V\, \text{d}\chi} 
  &= \frac{G_F^2 |V_{ub}^L|^2 m_B^3 }{ \pi^4} \bigg\{  ( \bar J_{1s}  \sin^2 \theta_V +  \bar J_{1c} \cos^2 \theta_V)
  - \frac13( \bar J_{2s} \sin^2 \theta_V + \bar J_{2c} \cos^2 \theta_V) \nonumber\\
&+ \frac23 \bar J_{3} \sin^2 \theta_V \, \cos 2 \chi + \frac{\pi}{4} \bar J_{5} \sin 2 \theta_V  \, \cos \chi
  + \frac{\pi}{4} \bar J_{7} \sin 2 \theta_V   \, \sin \chi + \frac23 \bar J_{9} \sin^2 \theta_V \, \sin 2 \chi \bigg\}\,,\\
\frac{\text{d}\Gamma}{\text{d}\cos\theta_\ell\, \text{d}\chi} 
  &= \frac{G_F^2 |V_{ub}^L|^2 m_B^3 }{3 \pi^4} \times \bigg\{ 2 \bar J_{1s}  + \bar J_{1c}
  + (2  \bar J_{2s}  +  \bar J_{2c} ) \cos 2\theta_\ell 
+ 2 \bar J_{3}  \sin^2 \theta_\ell \, \cos 2 \chi\nonumber\\
&+ ( 2 \bar J_{6s}  +  \bar J_{6c} ) \cos \theta_\ell + 2  \bar J_{9} \sin^2 \theta_\ell \, \sin 2 \chi
\bigg\}\,.
\end{align} 
Integrating over two angles, the rates become
\begin{align}
\frac{\text{d}\Gamma}{ \text{d}\cos\theta_\ell} 
  &= \frac{2 G_F^2 |V_{ub}^L|^2 m_B^3 }{3 \pi^3} \bigg\{ 2 \bar J_{1s}  + \bar J_{1c}
  (+ 2 \bar J_{2s} + \bar J_{2c})\cos 2\theta_\ell +( 2 \bar J_{6s}  + \bar J_{6c})\cos\theta_l \bigg\}\,, \\
\frac{\text{d}\Gamma}{\text{d}\cos\theta_V\,} 
  &= \frac{2 G_F^2 |V_{ub}^L|^2 m_B^3 }{ \pi^3} \bigg\{\bar J_{1s}  \sin^2 \theta_V + \bar J_{1c} \cos^2 \theta_V
  -\frac13 ( \bar J_{2s} \sin^2 \theta_V + \bar J_{2c} \cos^2 \theta_V) \bigg\}\,, \\
\frac{\text{d}\Gamma}{ \text{d}\chi} 
  &= \frac{G_F^2 |V_{ub}^L|^2 m_B^3 }{2 \pi^4}  \bigg\{ \frac83 \bar J_{1s}  + \frac43 \bar J_{1c} -  \frac89 \bar J_{2s}  + \frac49  \bar J_{2c}
  + \frac{16}{9}\bar J_{3}  \, \cos 2 \chi + \frac{16}{9} \bar J_{9} \, \sin 2 \chi \bigg\}\,.
\end{align}
\end{widetext}

\section{Branch Cut Uncertainty}
\label{bc_uncertainty}

Using $t\equiv q^2$, the bound on the
expansion coefficients of the form factors $A_l^X$ can be written as \cite{Bharucha:2010im}
\begin{equation}
  \frac{1}{2\pi} \int_0^{2\pi}\text{d}\phi \left| \Phi  A_l^X \right|^2 
  (e^{-i \phi}) \leq 1\,, \label{bound}
\end{equation}
where $A_l^X$ are projections onto the longitudinal $(l)$ and
transverse $(t)$ components,
    $|A_l^X (t)|^2 =  P_l^{\mu\nu} \langle \rho | j_\mu^X | B \rangle\langle B | j_\nu^X | \rho \rangle\,.$
For our model we need to redefine the kinematical factor with poles in the branch cut region into the matrix element  $|A_l^X (t)|^2 \to \frac{( t_- - t ) ( t_+ - t)}{3 t}
 |\widetilde A_l^X (t)|^2$.  Hence the $\Phi$ function is now given by $
\Phi_\text{new}(t) =  \sqrt{3 \frac{-t}{z(t,0)} \frac{t_- -t}{z(t,t_-)} (t - t_+ )}\Phi_{F}(t)$. Analyticity is restored by subtracting the branch cut
\begin{align}
  g(z) & \equiv \Phi(z)  \bar A_l^X(z) = \Phi_\text{new}(z) \widetilde A_l^X(z)\nonumber \\
  &\quad -\frac{1}{\pi}\int_{-1}^{z_\text{cut}} \text{d}x\, \frac{\Phi_\text{new}(x)\, \text{Im} F(x)}{x-z} \,.
\end{align}
The true analytic form factor is $\bar A_l^X(z)$, for which we can derive the
bound using $g(z)$. Here $z_\text{cut}\equiv z(t_\text{cut},t_0)$ is the
position of the lowest sub-threshold branch cut. We integrate only from $z\equiv
z(t_+,t_0)=-1 $, because everything above the two-particle threshold is being
taken care of already.  The function $\text{Im}\, F(x)$ is connected to the
branch cut, and no analytic expression of this exists. However we may model it
with an ansatz, since its origin is related to matrix elements of the form
$\text{Im}\,\,\langle 0 | j^X | B h \rangle$, where $h$ is a (combination) of
allowed light hadrons in the final state. This is an intermediate state of the
transition $B\to \rho$. The model function should fulfill the conditions (i)
vanish as $t\to \infty$ (or get constant for a finite $t$ interval), (ii) start
with zero at the threshold point, and (iii) it should be a contiguous function.
We will model this function inspired by the optical theorem and saturation of
the lowest states due to phase-space and multiplicity suppression. The first
factor is related to the ``production coupling'' of the state, and the second is
related to the kinematics of the branch cut.

We try a model function inspired by $e^+ e^- \to \mu^+ \mu^-$ scattering, which
gives larger contributions than, e.g., a model used in \cite{Caprini:1995wq}
\begin{equation}
    \text{Im} F(t) = C \sqrt{t_+} \sqrt{1-\frac{t_\text{cut}}{t}}
    \left(1+ \frac{t_\text{cut}}{2t} \right),\,\quad t_\text{cut} \leq t\leq t_+\,,
\end{equation}
where $C$ is in general a dimensionless quantity. The saturation of the lowest
state has been assumed in a dispersion relation as   $\text{Im}\,\,\langle 0 |
j^X | B h \rangle$ to estimate $C$~\cite{Caprini:1995wq}. Integrating over the
phase-space region in question, the authors have found a slow varying number of
order one.  Another possibility is to assume the on-shell production of the
leading branch-cut state out of the vacuum. We use a generic meson coupling
constant model in the narrow width approximation. We normalize the production
current to the threshold mass of the system over the width, similarly to an
intermediate on-shell state. Each additional particle has a phase-space
suppression factor of $1/(4\pi)^2$. We neglect further suppression by spin and
isospin quantum numbers, i.e., Clebsch-Gordan coefficients. We estimate the
coupling constant by the relation
\begin{align}
    C &\approx \frac{g_{B n \pi}^2 (m_B+ n \,m_\pi)}{\Gamma_{B \pi\pi} (16\pi^2)^{n-1}}\, c  \,\overset{\overset{\text{narrow}}{\text{\tiny width}}}{\underset{n=2}{\longrightarrow}}\, \frac{3}{\pi} \frac{1+\frac{2 m_\pi}{m_B}}{(1-\frac{4 m_\pi^2}{m_B^2})^\frac32 }\, c
\end{align}
with a dimensionless constant $c \sim {\cal O}(1)$. The coupling gets smaller for a higher multiplicity state, as expected, and we focus on the leading contribution $n=2$.

In some cases, there may be additional suppressions, e.g., for isospin violating
transitions or OZI-suppressed  decays, leading to a smaller coupling $C$. This
could happen in $B_c$ decays, however it is not present in $B\to \rho$. 

This most important contribution will be from $B+ 2 \pi$, with $n=2$, which is
neither isospin, nor spin nor OZI suppressed. Numerically we have
\begin{equation*}
    t_+ \approx  36.65\,\text{GeV}^2 \quad 
    t_\text{cut} \approx 30.79\,\text{GeV}^2 \quad 
    z_\text{cut} \approx -0.344\,.
\end{equation*}
For comparison the physical form factor values are
   $ z(0,t_0) \approx 0.10\,,\,  z(t_-,t_0) \approx -0.10\,.$
In summary we estimate in our approach $C\approx 1.01 c$, hence an order one number as   as in~\cite{Caprini:1995wq}. Subsequently we will assume a (hopefully) conservative estimate of $C\approx 10$.

For the estimate of this branch cut influence, we use the Minkowski inequality,
which states for (Lebesgue) integrable functions
    $||f+g||_p \leq ||f||_p+||+g||_p$
for any norm $p>1$, which is defined as
    $||f||_p = \left(\int \text{d}\mu | f | ^p \right)^\frac{1}{p}\,.$
Thus for $p=2$ in our case we can write the inequality \eqref{bound} as
\begin{align}
    &\sqrt{ \int_0^{2\pi}\text{d}\phi \left| \Phi  \widetilde A_l^X \right|^2 (e^{-i \phi})}
    \leq \sqrt{ \int_0^{2\pi}\text{d}\phi \left| g(e^{-i \phi})\right|^2 } \nonumber \\
    &\phantom{\leq\,}\,+\sqrt{ \int_0^{2\pi}\text{d}\phi \left|\frac{1}{\pi}\int_{-1}^{z_\text{cut}} \text{d}x\,\, \frac{\Phi(x)\, \text{Im} F(x)}{x-e^{-i \phi}}\right|^2}\nonumber\\
    &\leq \sqrt{2\pi }(1  + I_\text{cut})\,. \label{boundhelp}
\end{align}
We have used the fact, that the integral of the analytic function $g(z)$
fulfills  the bound, while the left-hand side is the ``true'' relation. Hence we
can estimate the deviation from the bound through this additional cut removal
function by the model and numerical integration. For the numerical evaluation we
take two subtractions $n=2$, as the contributing form factors require these
numbers of subtractions. Furthermore we need to use a numerical value for
$\chi(n=2)$. Since this has not been calculated by us, we take the value for the
transverse form factor part with two subtractions from \cite{Bharucha:2010im}
with $\chi(2) = 0.0116/m_b^2$. For this being a rough estimate of the
branch cut uncertainty, this value will be sufficient. 

In fact we derived an indirect correction to the bound, which constrains the
expansion parameters of the residual $q^2$ dependence. Our calculated form
factor dependence thus fulfills the corrected bound
\begin{align}
 \frac{1}{2\pi}\int_0^{2\pi}\text{d}\phi \left| \Phi  A_l^X \right|^2 (e^{-i \phi}) &= \frac{1}{B_{FF}\, \Phi_{FF}} \,
  \sum_{k=0}^K \alpha^{FF}_k\, z(q^2,\, q^2_0)^k \nonumber \\
    &\leq (1 + I_\text{cut})^2\,,
\end{align}
which has a correction term of the form $2 I_\text{cut}+ I_\text{cut}^2$. If
these additional terms are sufficiently small, the bound on the residual
dependence is not changed dramatically and from this we conclude the shape is
not changed by these branch cut singularities. Note especially that the leading
contribution to $\alpha_0$ is only a constant shift, and no shape distortion. A
modification of the first shape influencing expansion coefficient would be
multiplied by a number $|z| \lesssim 0.1$. Numerically this amounts to 
\begin{align}
    I_\text{cut} &= \sqrt{\frac{1}{2\pi} \int_0^{2\pi}\text{d}\phi \left|\frac{1}{\pi}\int_{-1}^{z_\text{cut}} \text{d}x\,\, \frac{\Phi(x)\, \text{Im} F(x)}{x-e^{-i \phi}}\right|^2}
    I_\text{cut} \approx 0.11 \,.
\end{align}
This now has to be compared with the bound 
\begin{equation}
 \frac{1}{2\pi }\int_0^{2\pi}\text{d}\phi \left| \Phi  A_l^X \right|^2 (e^{-i \phi}) \leq (1 + I_\text{cut})^2\,,
\end{equation}
At first sight, this seems to indicate an order ${\cal O}(10\%)$ correction to
the form factor bound itself. However, that will affect the bound on all
expansion coefficients, and does not mean a 10\% contribution to the leading
coefficient squared of all of the form factors. For the expansion parameter
fulfills $-0.1 \lesssim z \lesssim 0.1$, and so the larger allowance of higher
order coefficient contributions will not change the slope dramatically.
Furthermore it has to be compared with the ${\cal O}(10\%)$ uncertainty of each
form factor data point, which is of the same order of magnitude. Furthermore
this bound constraints a linear combination of all transverse form factors,
which can formally easy be derived in the helicity eigenbasis. So in total even
with this correction only a mild influence on the individual form factor shape
is expected. 

In Ref.~\cite{Caprini:1995wq} a similar model, which gives smaller numbers in
this case, was applied to the $B\to D^* \ell \bar\nu_\ell$ decay in a slightly
different approach. The have found a $10^{-3}\cdot C $ influence on the shape
parameters, which they dubbed to be very small. This is to be compared to our
numbers, which are a bit higher as expected but still ok.

Thus we conclude the influence of branch cuts is at most at the few percent
level on the form factor shape as well as the parametrization. Therefore
regarding our precision, this effect can be safely neglected.

\end{document}